\documentclass[preprint,12pt]{elsarticle}

\usepackage{color}
\usepackage{relsize}
\usepackage{graphicx}
\usepackage{amssymb}
\usepackage{subfig}
\usepackage{epstopdf}
\usepackage{multirow}
\usepackage{amsmath}
\usepackage{caption}
\usepackage{array}
\usepackage{url}
\usepackage{longtable}
\usepackage{tabularx}
\usepackage{booktabs}
\journal{Journal of Network and Computer Applications}
\date{}
\begin{document}
\setlength{\abovedisplayskip}{1pt}
\setlength{\belowdisplayskip}{1pt}
\parskip 0pt
\begin{frontmatter}



\title{Reliability Modeling and Analysis of Communication Networks}
\author[label1]{Waqar Ahmed\corref{cor1}}
\cortext[cor1]{Waqar Ahmed}
\address[label1]{School of Electrical Engineering and Computer Science\\
National University of Sciences and Technology, Islamabad, Pakistan }
\ead{waqar.ahmad@seecs.nust.edu.pk}
\author[label1]{Osman Hasan}
\ead{osman.hasan@seecs.nust.edu.pk}
\author[label1]{Usman Pervez}
\ead{usman.pervez@seecs.nust.edu.pk}
\author[label2]{Junaid Qadir}
\address[label2]{Information Technology University (ITU)-Punjab, Lahore, Pakistan}
\ead{junaid.qadir@itu.edu.pk}


\begin{abstract}
%% Text of abstrac
In recent times, the functioning of various aspects of modern society---ranging from the various infrastructural utilities such as electrical power, water to socio-economical aspects such as  telecommunications, business, commerce, education---has become critically reliant on communication networks, and particularly on the Internet. With the migration of critical facilities to the Internet, it has become vitally important to ensure the reliability and availability of networks. In this paper, we study various modeling and analysis techniques that can aid in the study of reliability of communication networks. In this regard, we provide background on the modeling techniques (such as reliability block diagrams, fault trees, Markov chains, etc.) and analysis techniques (such as mathematical analytical methods, simulation methods, and formal methods). Apart from providing the necessary background, we also critically evaluate the pros and cons of different approaches, and provide a detailed survey of their applications in communication networks. To the best of our knowledge, this is the first in-depth review of the application of reliability modeling and analysis techniques in communication networks.
\end{abstract}

\end{frontmatter}

\section{Introduction}
\label{sec:intro}

Communication networks have become an integral part of our daily life with applications ranging from ubiquitous hand held devices (like cell phones and remote car keys) to sophisticated equipment used in aircrafts, power systems, nuclear plants and healthcare devices. Given the safety and financial-critical nature of many of these applications, the failures in the underlying network elements can significantly affect the performance of network services and have detrimental, or even catastrophic, results.

For instance, the virtualized environment of a virtual data center in a cloud computing network  is often subjected to transient latency, dropped packets, and full-blown network partitions \cite{berger2009security}. A study for the cloud computing vulnerabilities shows that there were about 172 unique cloud computing outage incidents between 2008 and 2012 \cite{ko2013cloud}. The major causes of these incidents include (i) insecure interfaces and application programmer interfaces (APIs), (ii) data loss and leakages, and (iii) hardware failures. The main victims of these vulnerabilities include Google, Amazon, Microsoft and Apple and the vulnerabilities resulted in heavy financial losses \cite{ko2013cloud}. It is reported that the Amazon Web Service (AWS) suffered an unavailability for 12 hours, in April 21, 2011, causing hundreds of high-profile Web sites to go offline \cite{bailis2014network}, which resulted in a loss of 66,240 US\$ per minute downtime of its services. In order to predict such problems beforehand, several reliability modeling and analysis  techniques are utilized  \cite{jereb1998network}.
 %{\color{red} Write a powerful first paragraph. Give some high-profile failure examples from the article `the network is reliable' published in ACM Queue \cite{bailis2014network} and the resulting losses, thereby motivating dependability modeling and analysis.}

\begin{figure}
  \centering
  \subfloat[]{\includegraphics[width=0.45\textwidth,trim = {2cm 0cm 5cm 0cm},clip,keepaspectratio]{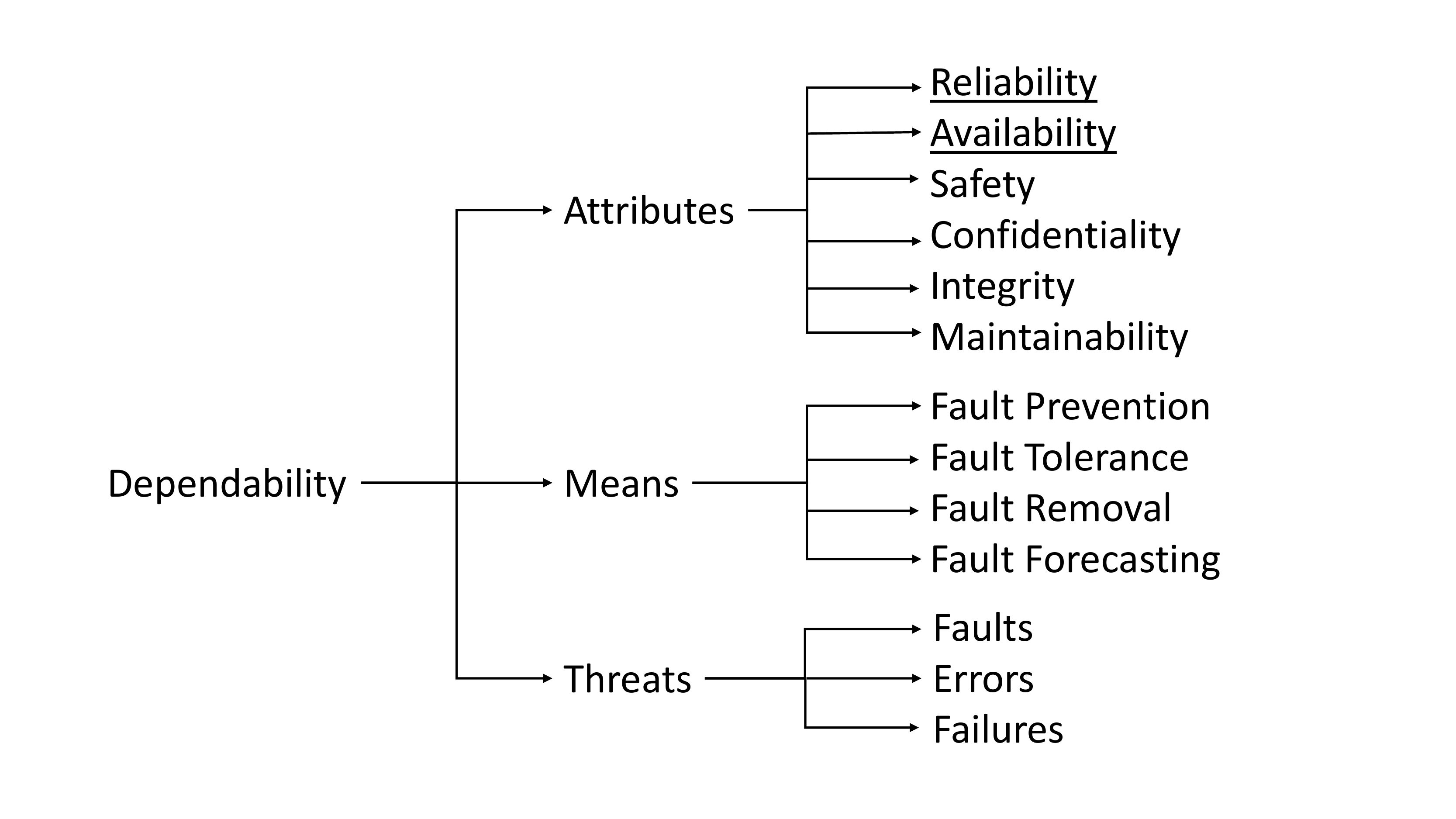}}
  \subfloat[]{ \includegraphics[width=0.55\textwidth, keepaspectratio]{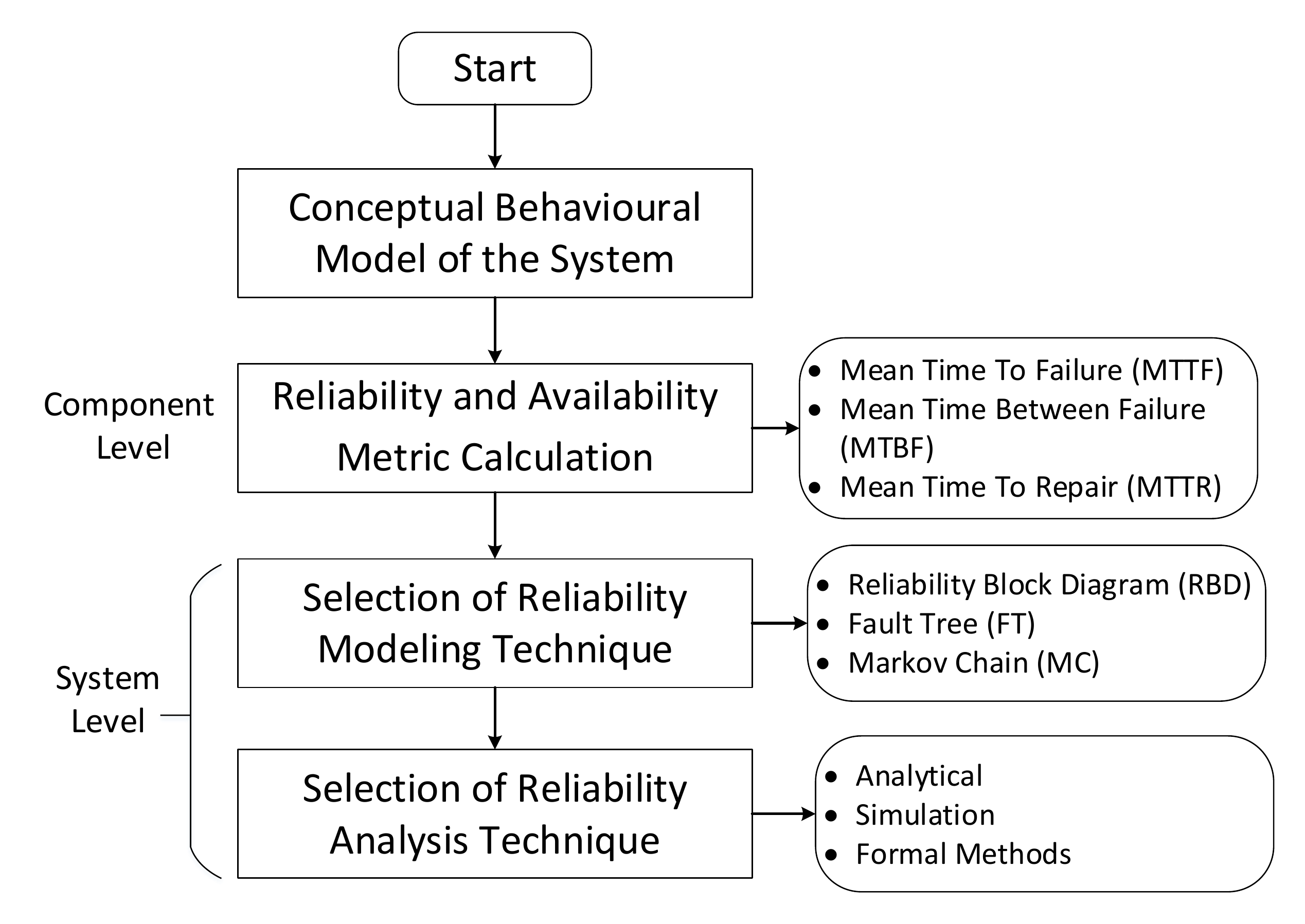}}
  \caption{(a) Taxonomy of Dependability Attributes (b) Steps for Reliability Assessment}\label{Dep_dia}
\end{figure}

\textit{Reliability} is defined as the probability of a system or a sub-component functioning correctly under certain conditions over a specified interval of time \cite{villemeur1992reliability}. For instance, the reliability of network nodes, termed as the \textit{terminal reliability}, is the probability that a set of operational edges provides communication paths between every pair of nodes \cite{altiparmak2009general}. Another closely related concept with reliability is \textit{availability} \cite{al2009comparative}, which can be defined as the probability that a component will be available when demanded \cite{avizienis2001fundamental}. As an example, the availability of a mesh network is the probability that every mesh node is connected to at least one gateway \cite{pathak2013designing}. To understand the difference between the reliability and availability concepts, it is important to realize that reliability refers to failure-free operation during an interval, while availability refers to failure-free operation at a given instant of time \cite{pradhan1996fault,trivedi2008probability}. The availability of a system is typically measured as a function of reliability and \textit{maintainability}, which is defined as the probability of performing a successful repair action of a system under a given time and stated conditions \cite{villemeur1992reliability}. Additionally, if we keep the maintainability measure constant, the availability of the system is directly proportional to the reliability of the system, i.e., an increase in reliability increases the availability of the system and vice versa \cite{Weibull_15}. \textit{Dependability} \cite{al2009comparative}, i.e., an umbrella concept that subsumes reliability  and availability  considerations,  is primarily defined as the ability of a system to perform its desired function or tasks faultlessly in a certain environment on a planned time period \cite{laprie1992dependability}. Many authors describe dependability of a system as a set of properties or attributes such as reliability, maintainability, safety, availability, confidentiality, and integrity \cite{edwards1994building,avizienis2001fundamental,kyriakopoulos2000dependability}. Some of these attributes, such as reliability and availability, are quantitative whereas some are qualitative, for instance, safety \cite{al2009comparative}. A generalized view of dependability attributes along with its threats and the means to achieve dependability are shown in Figure \ref{Dep_dia}. This paper is mainly focused on reliability and availability because of its importance and wide utilization in the area of communication networks.

Reliability analysis plays a vital role in the identification of existing problems in communication networks, prevention of future problems by improving network design, prediction of the behavior of telecommunication networks versus time and providing decision making support in designing performance efficient telecommunication networks \cite{bernardi2013dependability}. In particular, the reliability prediction allow us to determine the redundancy requirements in the given network, the ability of a network to maintain an acceptable reliability level under extreme environmental conditions and assess the impact of design changes on the reliability of the overall network \cite{misra1992reliability}. Similarly, the availability prediction allows us to evaluate various maintenance design options to achieve the desired availability of the network \cite{spragins1981communication}.

The qualitative and quantitative reliability analysis requires the selection of an appropriate mathematical modeling and analysis technique. The modeling technique must be able to effectively capture the important parameters of the real system and the analysis technique should be capable of providing insights into the system behavior without running (or executing) the real system. There are numerous techniques that can provide analysis in the \textit{early} phase, when only initial details are available of the design, and there are other techniques that cater the analysis of the \textit{later} design phases, when more precise implementation details are available \cite{bernardi2013dependability}. Some of the most widely used formalisms/techniques for modeling reliability and availability of communication networks are: Reliability Block Diagram (RBD) \cite{vcepin2011reliability}, Fault Tree (FT) \cite{vesely1981fault}, and Markov Chain (MC) \cite{gilks2005Markov}. Traditionally, the models developed using these techniques are analyzed using analytical methods or simulation tools.

Recently, the computer networks community has shown great interest in the utilization of formal methods \cite{qadir2015applying} as a reliability analysis technique for communication networks. Formal methods use mathematical logic to precisely model the system's intended behavior and deploy mathematical reasoning to construct an irrefutable proof that the given system satisfies its requirements. This kind of mathematical modeling and analysis makes formal methods an accurate and rigorous analysis method compared to the traditional analytical and simulation based analysis. This is because the involvement of manual manipulation/simplification in paper-and-pencil based analytical methods makes the analysis error-prone, especially while analyzing large systems. In addition, the key assumptions for analytical proofs may be implicit in the mind of the mathematician and not documented, which can create problems when the system is designed and implemented.

An overview of the essential steps for conducting reliability assessment of communication networks, is shown in Figure \ref{Dep_dia}(b). The main steps are: (i) Development a conceptual behavioral model of the given system; (ii) Calculation of reliability and availability metrics; (iii) Selection of reliability modeling techniques; and (iv) Selection of reliability analysis techniques. The first step of the reliability assessment starts with the construction of the conceptual behavioral model of the system. In this step, the network design engineers describe the intended mode of communication and the desired network behaviours, such as network protocols, network topologies and fault tolerance, of the given communication network. The second step is the calculation of basic metrics of reliability and availability, such as \textit{mean-time to failure} (MTTF) \cite{stanley2011mtbf}, \textit{mean-time between failure} (MTBF) \cite{speaks2010reliability} and \textit{mean-time to repair} (MTTR) \cite{stanley2011mtbf}, at the individual \textit{component level} of the communication network. The metrics of MTTF and MTBF are the basic metrics of reliability, usually measured in the unit of hours for non-repairable and repairable systems, respectively. For example, these metrics can be obtained by statistically calculating the failure rates of the routers that are generally modeled as the network nodes.  Suppose 10 routers are tested for 500 hours and $2$ failures occur during this time. We can then estimate the failure rate of the routers by using $\lambda = \frac{number \ of \ failures}{number \ of \ routers * Total \ time} = 0.004 \ failure/hour$ and $\mathit{\frac{1}{\lambda} = 2500 \ hours/failure }$. As mentioned earlier, the availability is estimated by using the reliability and the maintainability metrics, i.e., the availability of each network component is based on the MTBF and MTTR measures of each network component \cite{schneidewind2012computer}. So, the availability is calculated from reliability and also incorporate repair. The typical formula for availability calculation is $\frac{MTTF}{MTTR + MTTF}$.

The details of the above-mentioned metrics, at the component level, are essentially required by the reliability modeling techniques to estimate the reliability at the system level. The next step is the selection of an appropriate reliability modeling technique, such as RBD, FT or MC. In some cases, the selection of modeling technique is a direct consequence of the decisions taken in the first step. For instance, Markov chains could be the obvious choice if the system is dynamic in nature, otherwise, RBD and FT are generally the first choice of network design engineers due to their ability to model complex network systems in a simple understandable way.  However, it has been observed from the literature that some network systems can be easily modeled by more than one reliability modeling techniques. In this situation, the network design engineers articulate their computational resources and choose the reliability analysis technique accordingly, at the last step. For instance, if the state-space modeling techniques are selected then analytical techniques can be used if the given system is not too large. Simulation techniques, on the other hand, are well suited for larger systems but for safety-critical network systems, state-space base formal methods, like Perti Nets and Model checking, are becoming a common choice of network design engineers.

To understand the above-mentioned reliability assessment steps, we consider a network model, which is made up of $5$ wireless nodes. Assuming that as soon as $3$ of these nodes are out-of-service then the communication generally degrades so much that the system can be considered as unavailable. But, it may happen that the traffic is so low that a degraded level of transmission can also become acceptable. The reliability analysis of this Wireless sensor network (WSN) allows us to find quantitative information about its availability. The first step is the development of a conceptual model of the network behaviour and it usually requires a considerable amount of effort. In the literature, the reliability analysis of similar kinds of WSN systems has been studied in detail \cite{silva2012reliability,bruneo2010dependability}.  The authors, in \cite{bruneo2010dependability}, defined reliability of a WSN system as a probability that at least $k$ sensor nodes are alive at time $t$. Then they analyzed the reliability and availability of WSN systems by using network topologies, such as star, tree and mesh, and derived the analytical relationships to determine the least number of nodes that are essentially required for the network to be available. The second step, in the above example, is to determine the MTTF and MTTR of the WSN nodes \cite{silva2012reliability}. In the third step, we can choose a modeling method, such as a Markov chain, for the availability analysis of WSN \cite{bruneo2010dependability}. The exponential distributions can be used to model the failure characteristics of the WSN nodes. Lastly, the node availability expressions of the WSNs can be analytically derived by using the Chapman-Kolmogorov equations in order to determine the producibility of a WSN, which is the probability that at least $k$ of the $n$ nodes are
able to send data to the sink at time $t$ \cite{bruneo2010dependability}. Simulations can also be used to analyze the behaviour of WSN producibility for different parameter values.

\subsection{Contributions and Organization of this Paper}
\label{sec:contribution}

The purpose of this survey paper is to provide a generic overview of the major reliability modeling and analysis techniques in the domain of communication networks. Based on the three kind of reliability analysis techniques, we divide this survey into three broad parts: (i) analytical (ii) simulation tools and (iii) formal methods. Each of these parts encompasses a comprehensive survey of using the underlying analysis technique with the three reliability modeling techniques: RBDs, FTs, and MCs, in the context of communication networks. The main focus of the paper is to study the utilization of these modeling and analysis techniques in the domain of telecommunication networks and thus gain insights about the strengths and weaknesses of these methods and how to use them in the most effective manner.

It is important to note that the paper is unique compared to the existing surveys and tutorials on reliability analysis \cite{trivedi1993reliability,bernardi2012dependability,venkatesan2013survey,al2009comparative} due to its exclusive focus on reliability modeling and analysis in communication networks. Moreover, to the best of our knowledge, there is no study available that characterizes the reliability analysis methods and also considers the recent trend of using formal methods for reliability analysis. A discussion about the pros and cons of reliability analysis and modeling techniques is also a prominent aspect of this work.

The rest of the paper is organized as follows: Sections \ref{sec:depend_model} and \ref{sec:depend_analysis} describe the reliability models and their corresponding analysis methods, respectively. Sections \ref{sec:AnalysisWithRBD}, \ref{sec:ReliabilityFaultTree} and \ref{sec:ReliabilityMC} present the survey of reliability modeling and analysis of communication network domain using RBD, FT, and MC, respectively. In Section \ref{sec:InsightsPitfalls}, we provide a graphical overview of the survey paper by using a milestone timeline graph and an histogram plot. This section also presents insights and critical analysis of the pros and cons of the various reliability modeling and analysis approaches. Finally, the conclusion is presented in Section \ref{sec:conclusions}. For better readability, the acronyms used in this paper are collected in Table \ref{tab:acronyms} as a convenient reference.

\begin{table}[!ht]
\caption{Acronyms used in this paper.}
\label{tab:acronyms}
\footnotesize
\centering
\scalebox{0.9}{
\begin{tabular}{p{1.7cm}p{5cm}p{1.7cm}p{5cm}}
\toprule
\textbf{\textit{Acronym}} & \textbf{\textit{Expanded Form}} & \textbf{\textit{Acronym}} & \textbf{\textit{Expanded Form}}\\
\midrule
3RIS &  Resiliency, Reliability, Redundancy by Infrastructure Sharing & MRM  & Markov Reward Modeling\\
B-ISDN & Broadband Integrated Service Network & OLSR &  Optimized Link State Routing\\
CBTC & Communication-Based Train Control System & PON  &  Passive Optical Network\\
CCS  &  Communication Control System & RBD  &  Reliability Block Diagram\\
CTMC & Continuous Time Markov Chain & RPM  &  Randomised Pulse Modulation\\
DCN  &  Data Communication System & SAS  &  Substation Automation System\\
DRBD &  Dynamic Reliability Block Diagram & SCADA & Supervision, Control and Data Acquisition System\\
DTMC & Discrete Time Markov Chain & SEN &  Shuffle Exchange Networks\\
FASP &  Fast and Secure Protocol & SNMP &  Simple Network Management Protocol\\
FDDI &  Fiber Distributed-Data Interface & SPC  &  System Protection Center\\
FT   &  Fault Tree & SPN  &  Stochastic Perti Nets\\
GPS  &  Global Positioning System & SRN  &  Stochastic Reward Nets\\
GSPN &  General Stochastic Petri Nets & TLR  &  Trust Levels Routing Protocol\\
IABN &  Irregular Augmented Baseline Network & VANETs & Vehicular ad-hoc Networks\\
ICT  &  Information Communication Technology & VoIP &  Voice over Internet Protocol\\
LAN  &  Local Area Network & WAPS &  Wide Area Protection Communication System\\
LEO & Low Earth Orbit & WLAN &  Wireless Local Area Network\\
LPC  &  Regional Protection Center & WSN &  Wireless Sensor Network\\
MC   &  Markov Chain & &\\
MINs  &  Multistage Interconnection Networks & &\\
MAS  &  Mission Avionics System & &\\
\bottomrule
\end{tabular}}
\end{table}

\section{Reliability Models}
\label{sec:depend_model}

Reliability assessment techniques can be utilized in every design phase of the system or component including development, operation and maintenance. FT and RBD based models are usually used to provide reliability and availability estimates for both \textit{early} and \textit{later} stages of the design, where the system models are more refined and have more detailed specifications compared to the the early stage system models \cite{Dependability2013tech}. While on the other hand, Markov chain based models are mainly used in the \textit{later} design phase to perform trade-off analysis among different design alternatives when the detailed specification of the design becomes available. In addition, when the system is deployed, these modeling techniques can be beneficial in order to estimate the frequency of maintenance and part replacement in the design, which allows us to determine the life cost of the system elements or components.

This section provides a brief description about the reliability models, i.e., Reliability Block Diagram, Fault Trees, Markov Chains and Bayesian Networks. These models provide basis for conducting the reliability analysis by using any one of the analysis approaches (analytical methods using paper-and-pencil based proofs, computer simulations or formal methods).

\subsection{Reliability Block Diagrams}
\label{sec:RBD}

\begin{figure}[!ht]
\centering \vspace{3pt}
\subfloat[]{\includegraphics[width=0.4\textwidth,keepaspectratio]{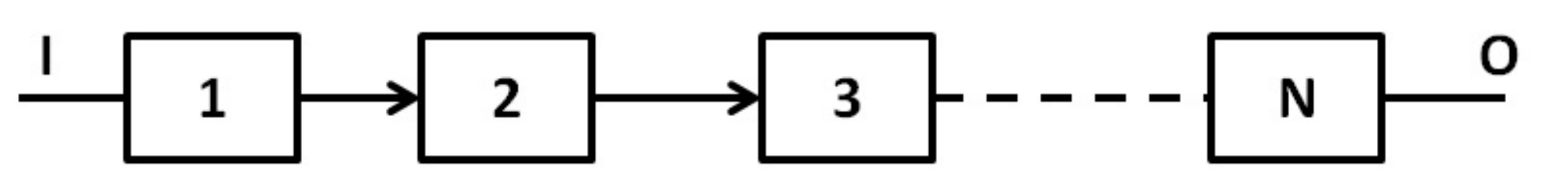}}
\subfloat[]{\includegraphics[width=0.5\textwidth, height= 3cm, keepaspectratio]{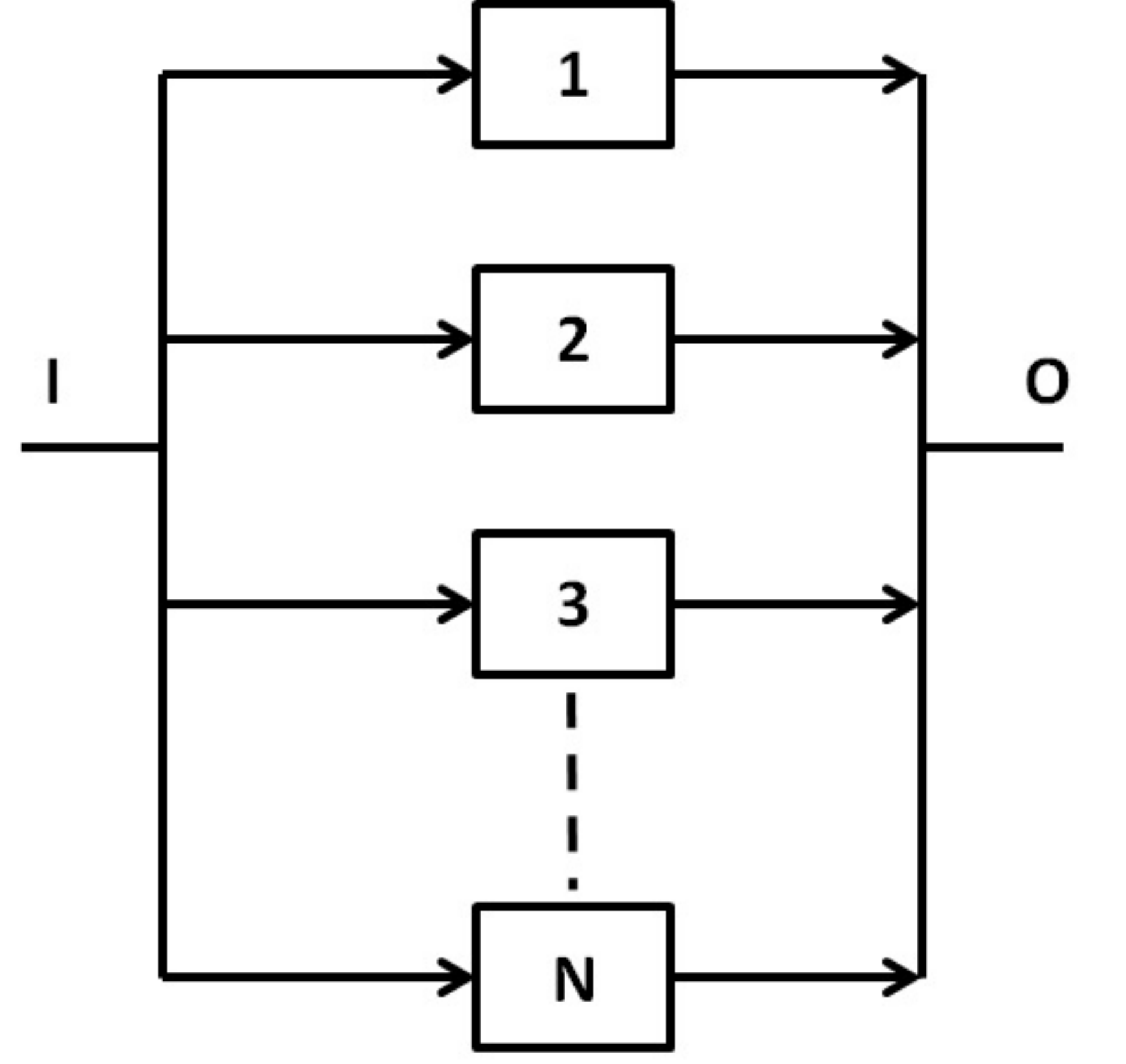}}\newline
\subfloat[]{\includegraphics[width=0.3\textwidth,height= 3cm, keepaspectratio]{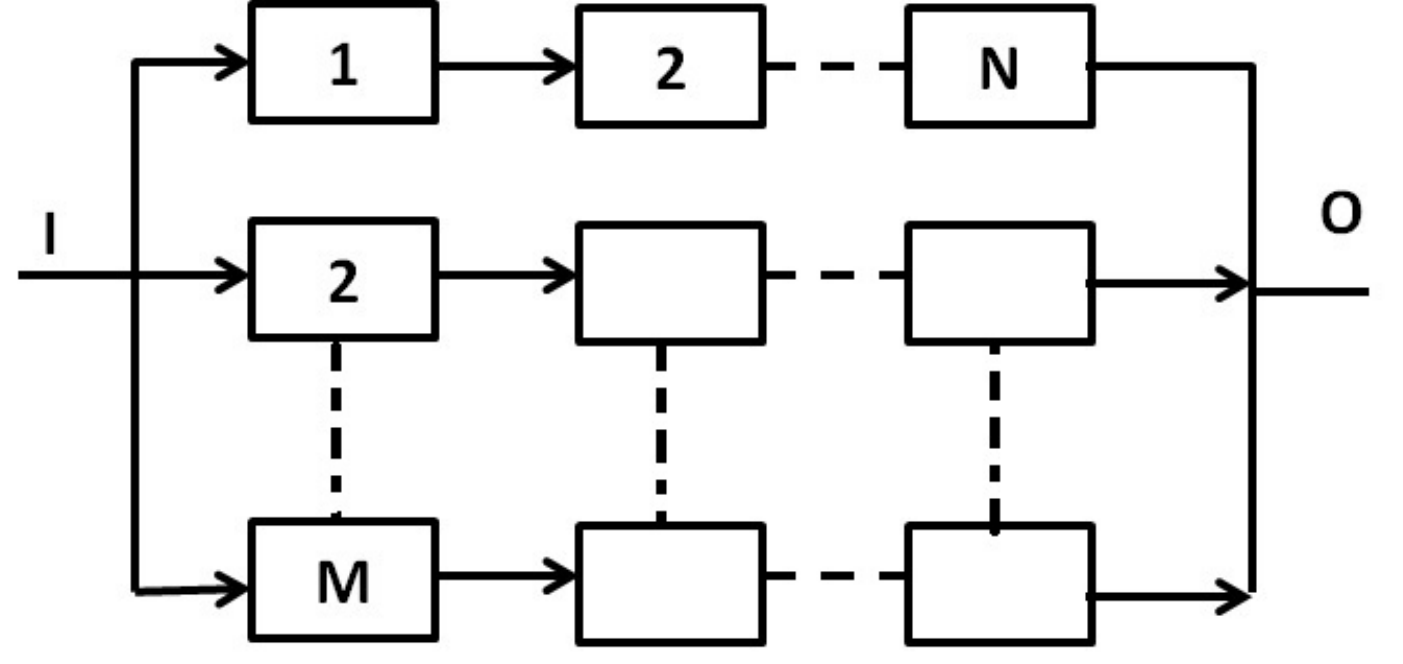}}
\subfloat[]{\includegraphics[width=0.4\textwidth,height= 3cm, keepaspectratio]{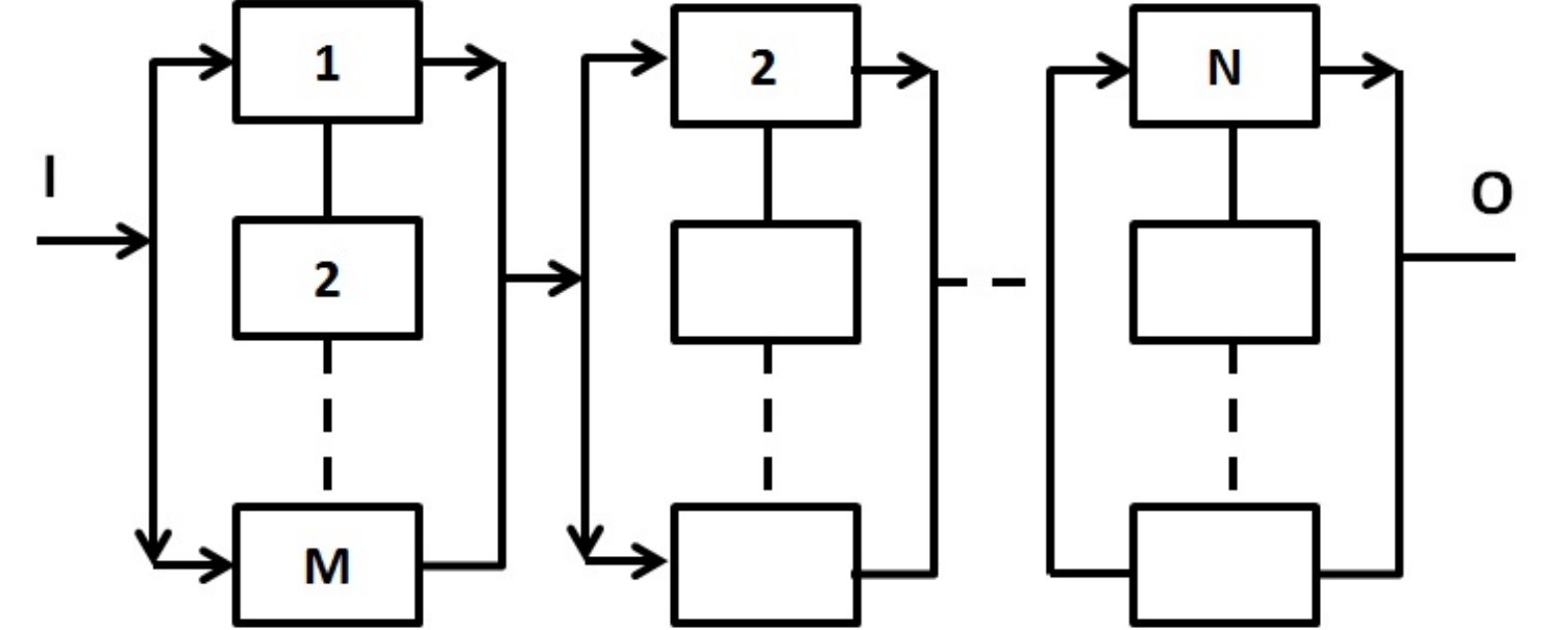}}\newline
\subfloat[]{\includegraphics[width=0.4\textwidth,height= 3cm, keepaspectratio]{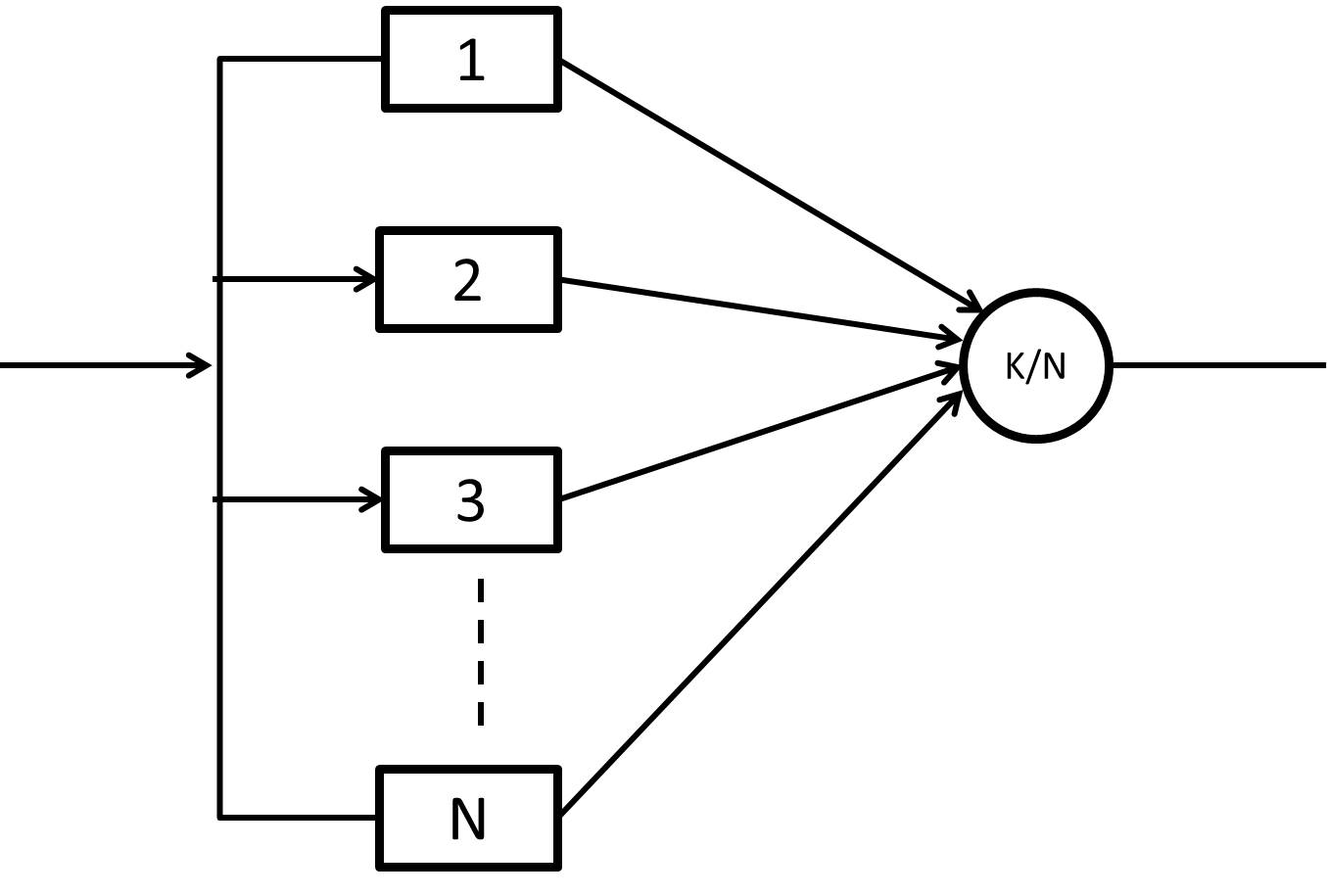}}
\caption{Reliability Block Diagram Configurations (a) Series (b) Parallel (c) Parallel-Series (d) Series-Parallel (e) \textit{k}-out-of-\textit{n}}
\label{fig:RBDs}
\end {figure}

Reliability Block Diagrams (RBDs) \cite{Bilinton_1992}, depicted in Figure \ref{fig:RBDs}, are graphical structures consisting of blocks and connector lines. The blocks usually represent the system components and the connection of these components is described by the connector lines. The input is given at one end of the RBD and the output is observed at the other end. The system is functional, if at least one path of properly functional components from input to output exists otherwise it fails.

An RBD construction can follow any of these three basic patterns of component connections: (i) series (ii) active redundancy or (iii) standby redundancy. In the series connection, shown in Figure \ref{fig:RBDs}(a), all the components should be functional for the system to be remain functional. Whereas, in active redundancy all the component in the redundant stages must be in active state. The components in active redundancy, shown in Figure \ref{fig:RBDs}(b), might be connected in a parallel structure (Figure \ref{fig:RBDs}(b)) or a combination of series and parallel structures as shown in Figures \ref{fig:RBDs}(c) and 3(d). In standby redundancy, not all the components are required to be active. A \textit{k}-out-of-\textit{n} (Figure \ref{fig:RBDs}(e)) is an example of standby redundancy, in which only \textit{k} components from \textit{n} components are supposed to be functional. The remaining components are in the standby mode and can be made useful, if required.

\subsection{Fault Tree}
\label{sec:FT}
\begin{figure}
  \centering
 \includegraphics[width=0.6\textwidth, height= 7cm, keepaspectratio]{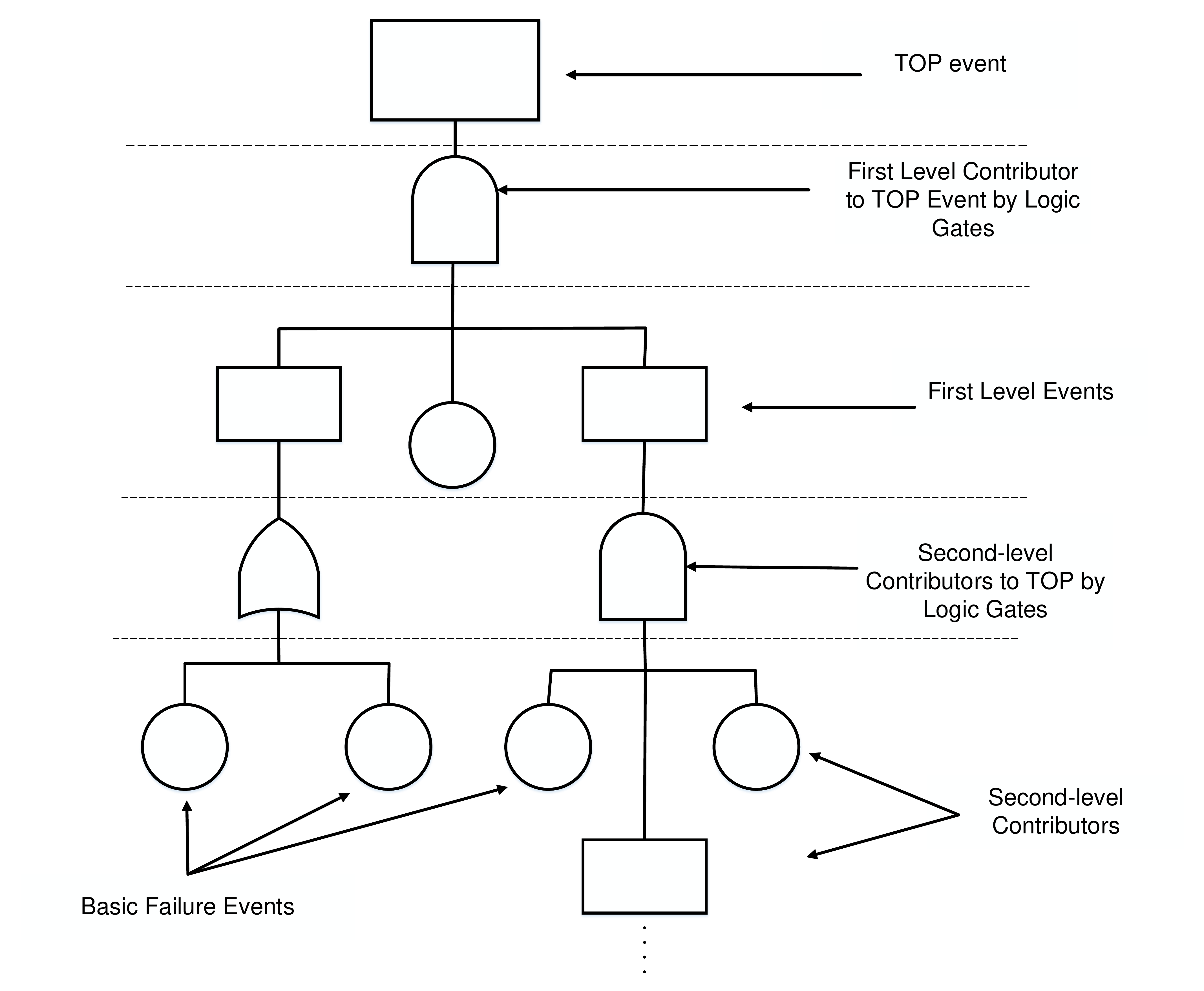}
  \caption{A Typical Fault Tree Diagram}\label{Fault_tree_pic}
\end{figure}

Fault Tree (FT ) \cite{ftta61025} is a graphical technique for analyzing the conditions and the factors causing an undesired \textit{top event}, i.e., a critical event, which can cause the whole system failure upon its occurrence. These causes of system failure are represented in the form of a tree rooted by the \textit{top event} as depicted in Figure \ref{Fault_tree_pic}. The preceding nodes of the fault tree are represented by \textit{gates}, which are used to link two or more \textit{cause events} causing one fault in a prescribed manner. For example, an OR FT gate can be used when one fault suffices to enforce the fault. On the other hand, the AND FT gate is used when all the cause events are essential for enforcing the fault. Besides these gates, there are some other gates, such as exclusive OR FT gate, priority FT gate and inhibit FT gate, which can be used to model the occurrence of faults due to the corresponding cause events \cite{ftta61025}.

Once the fault tree model is constructed, both qualitative and quantitative analysis can be carried out. A qualitative analysis in this context allows the identification of all combinations of basic failure events, known as cut sets, which can cause the top event to occur. The \textit{minimal} cut sets (MCS) are those cut sets that do not contain any subset of the basic cause events that are still a \textit{cut set} and are obtained by applying boolean algebraic operations on these cut sets. The smaller the number of basic cause events in these cut set, the modeled system is considered to be more resilient to failures. The quantitative analysis is used to evaluate the probability of occurrence of top event by considering these minimal cut sets which significantly contribute to the system failures.

The Dynamic Fault Tree (DFT) \cite{durga2009dynamic} is a type of FT that utilizes the time-variant FT gates, such as Priority AND (PAND), the sequence enforcing (SEQ), the standby or spare (SPARE), and the functional dependency (FDEP). These gates extend the traditional FT gates functionality as they can be used to model the dynamic behavior as well as the order of failure in a given system. For example, the PAND FT gate is considered to be at the failure state when all of its input components fail in a pre-assigned order (not forcibly), i.e., failure occurs from left to right in graphical notion. The SEQ FT gate forces the failure in a pre-assigned order and the SPARE FT gate models the substitution of one or more principle components with that of spare components having the same functionality \cite{durga2009dynamic}.

\subsection{Markov Chain}
\label{sec:MC}
\begin{figure}[!h]
  \centering
  \includegraphics[width=0.4\textwidth,height=5cm,keepaspectratio]{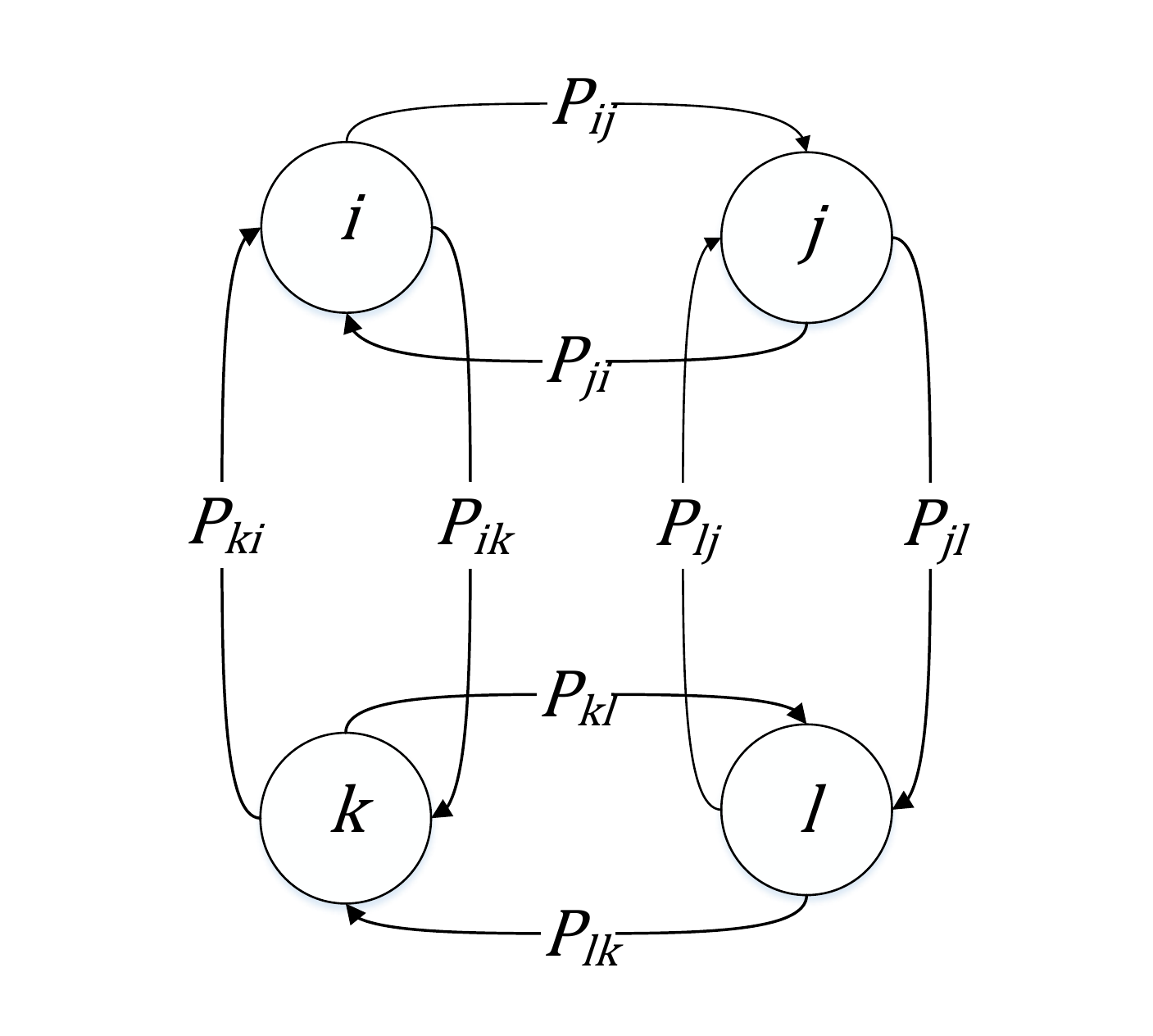}\\
  \caption{A Discrete Time Markov Chain Example}\label{DTMC_pic}
\end{figure}

 A Markov chain \cite{fugua2003applicability} is a stochastic process that consists of a set of states, i.e., $S =\{s_{0},s_{1},...,s_{n}\}$, and arcs, which are used to point the transition from one state to another. The initial state $s_{ini}$ and the probability $p_{ij}$ represent the starting state and the transition probability from state $s_{i}$ to state $s_{j}$, respectively. As shown in the example of Figure \ref{DTMC_pic}, the process starts from an initial state and transitions from the current state to the next state occur on the basis of transition probabilities, which only depend upon the current state. This provides the basis for Markov model and it is also known as the Markov or the memoryless property.

Markov chains are usually classified into two categories: Discrete Time Markov Chains (DTMC) and Continuous Time Markov Chains (CTMC) \cite{bhat1972elements}. Markovian models are frequently utilized for reliability analysis in scenarios where failure or repair events can occur at any point in time \cite{fugua2003applicability}. Similarly, the semi-Markov model \cite{walks1999semi}, which involves the concept of state and state transition, has also been used for the reliability analysis of the systems. The distinguishing feature of the semi-Markov models, compared to the Markov models, is that the transitions and the probability distributions depend on the time spent by a system at its present state \cite{fugua2003applicability}. In other words, the transition from the current state to the next state not only depends upon the transition probability but also the holding time, which is the time spent by a system at a current state. Moreover, the transition probabilities in a semi-Markov model can be non-exponential \cite{fugua2003applicability}.

Markov modeling has also been utilized for analyzing the \textit{dynamic} behavior of the other reliability models, i.e., RBD and FT. The notion of dynamic behavior, for reliability analysis, represents the evolution of system topology/configuration with respect of time. In case of the Dynamic Reliability Block Diagram (DRBD) \cite{distefano2006new}, the system is modeled in terms of \textit{states} of the component and the evolution of these components states is carried out by a sequence of \textit{events} \cite{distefano2006new}. A typical DRBD contains the following states: (i) \textit{Active}: the state of proper functioning of the component (ii) \textit{Failed}: the failure state of the component (iii) \textit{Standby}: the state depicting the case when the component is not in functional or in active condition but it can be activated. In addition, there are other states such as \textit{Hot}, \textit{Warm} and \textit{Cold}, representing the conditions when the system or component is disabled but energized, partially disabled and completely disabled, respectively. Also, there are some basic types of events for DRBD like \textit{Failure}, \textit{Wake-up}, \textit{Sleep} and \textit{Standby} switch, representing the events from active to failure state, standby to active state, active to standby state and transition between two standby states, respectively \cite{distefano2006new}.

\subsection{Bayesian Network Models}
\label{sec:BN}

\begin{figure}[!h]
  \centering
  \includegraphics[width=0.4\textwidth,height=5cm,keepaspectratio]{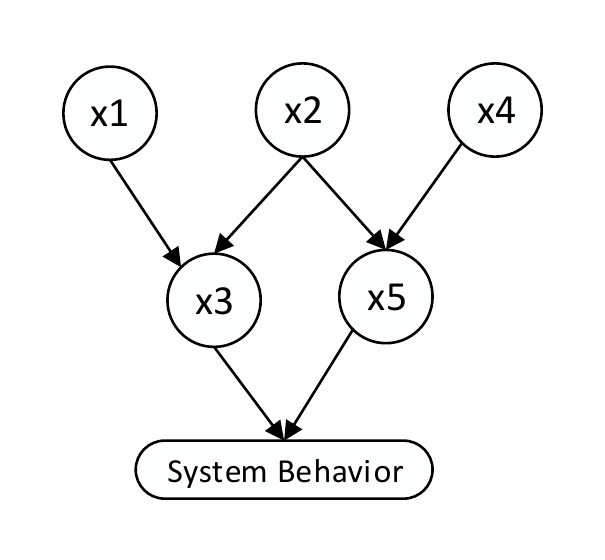}\\
  \caption{A Sample Bayesian Network }\label{BN_network}
\end{figure}

RBD and FT models fail to describe the system behavior when the component interaction of a system cannot be precisely defined.  Bayesian networks (BN) \cite{friedman1997Bayesian} allow us to cater for this problem by defining the component interactions probabilistically. Mathematically, a BN is a directed acyclic graph where the nodes represent a variable and the edges between the nodes represent the casual relationship between the nodes, as shown in Figure \ref{BN_network}. The top most nodes, X1, X2 and X4 representing the system components 1, 2 and 4, respectively, do not have any incoming edges therefore, they are conditionally independent of the rest of the components in the system. The probabilities that are assigned to these nodes should be known beforehand. This information can either be obtained with the help of a domain expert or using existing statistical data about the system. In the reliability analysis prospective, the variables in a BN are defined to represent system components and the edges or links represent the component interactions, which lead to the system failure or success \cite{doguc2009generic}.

\section{Reliability Analysis Techniques}
\label{sec:depend_analysis}

Traditionally, reliability analysis has been done using analytical approaches based on the paper-and-pencil proof methods and sampling-based computer simulations. However, these methods cannot ascertain absolute correctness due to their inherent limitations of incompleteness and error-prone nature. Formal methods have been recently proposed for the reliability analysis as well. However, these methods have a limited scope and thus cannot be used to analyze all kinds of complex engineering systems. In this section, we provide a brief overview about the above-mentioned reliability analysis techniques. Based on this description, a concise comparison between them is also presented in Section 7.

\subsection{Analytical}
\label{sec:analytical}

The main idea behind this approach is to analytically verify generic expressions for the reliability of the given system based on paper-and-pencil based proofs conducted by humans. Each reliability modeling technique, explained in Section \ref{sec:depend_model}, can be analytically analyzed to obtain equivalent mathematical expressions, which can be specialized to conduct the reliability analysis of the systems.

\subsubsection{Reliability Block Diagram Analysis}
\label{sec:RBD_anal}

RBD analysis can be used to conduct both qualitative and quantitative analysis of the systems. The former deals with the identification of critical or weak components of the given system. While in the latter, the reliability of the overall system is calculated on the basis of reliability of the individual components. All RBD configurations, shown in Figure \ref{fig:RBDs}, have an associated standard mathematical reliability equation which can be easily manipulated for conducting reliability analysis of large systems. Following are the reliability expressions for series, parallel, parallel-series, series-parallel and k-out-of-n RBD configurations, respectively:

\begin{equation}\label{eq1:series}
\small{R_{series}(t) = Pr (\bigcap_{i=1}^{N} A_{i}(t) )  = \prod_{i=1}^{N}R_{i}(t)}
 \end{equation}

\begin{equation}\label{eq2:parallel}
      \small{R_{parallel}(t) = Pr (\bigcup_{i=1}^{N}A_{i}(t) )= 1 - \prod_{i=1}^{N}(1 - R_{i}(t))}
      \end{equation}
\begin{equation}\label{eq3:parallel_series}
\small{R_{parallel-series} = Pr (\bigcup_{i=1}^{M} \bigcap_{j=1}^{N} A_{ij}(t))}
= 1- \prod_{i=1}^{M}(1 - \prod_{j=1}^{N} (R_{ij}(t)))
\end{equation}
\begin{equation}\label{eq4:series_parallel}
\begin{split}
  \small{R_{series-parallel}} & = \small{Pr (\bigcap_{i=1}^{N} \bigcup_{j=1}^{M} A_{ij}(t))}
     = \small{\prod_{i=1}^{N}(1 - \prod_{j=1}^{M} (1- R_{ij}(t)))}
\end{split}
\end{equation}

\begin{equation}\label{eq:k-out-nRBD}
\begin{split}
R_{k|n}(t) & = Pr (\cup_{i=k}^{n}\{\textit{exactly \textit{i} components functioning}\}) \\
& = \mathlarger{‎‎\Sigma}_{i=k}^{n} (\dbinom{n}{k} R^{i} (1 - R)^{n -1})
\end{split}
\end{equation}

\noindent where $A_{i}(t)$ represents the reliability event associated with the $i^{th}$ sub-component of the system, $R_{i}(t)$ and $R_{ij}(t)$ represent the reliability of the $i^{th}$ sub-component connected in series or parallel RBD configuration and the reliability of $j^{th}$ sub-component connected in the $i^{th}$ level of series-parallel or parallel-series RBD configuration, respectively. The symbol \textit{R} in Equation \ref{eq:k-out-nRBD} represents the reliability of the identical components that are connected in the k-out-of-n RBD structure. These equations (Eq. \ref{eq1:series}--\ref{eq:k-out-nRBD}) can be utilized for availability analysis by replacing the reliability event with the availability event and vice versa \cite{Matos2014}.

\begin{figure}[!h]
  \centering
  \includegraphics[width=0.7\textwidth]{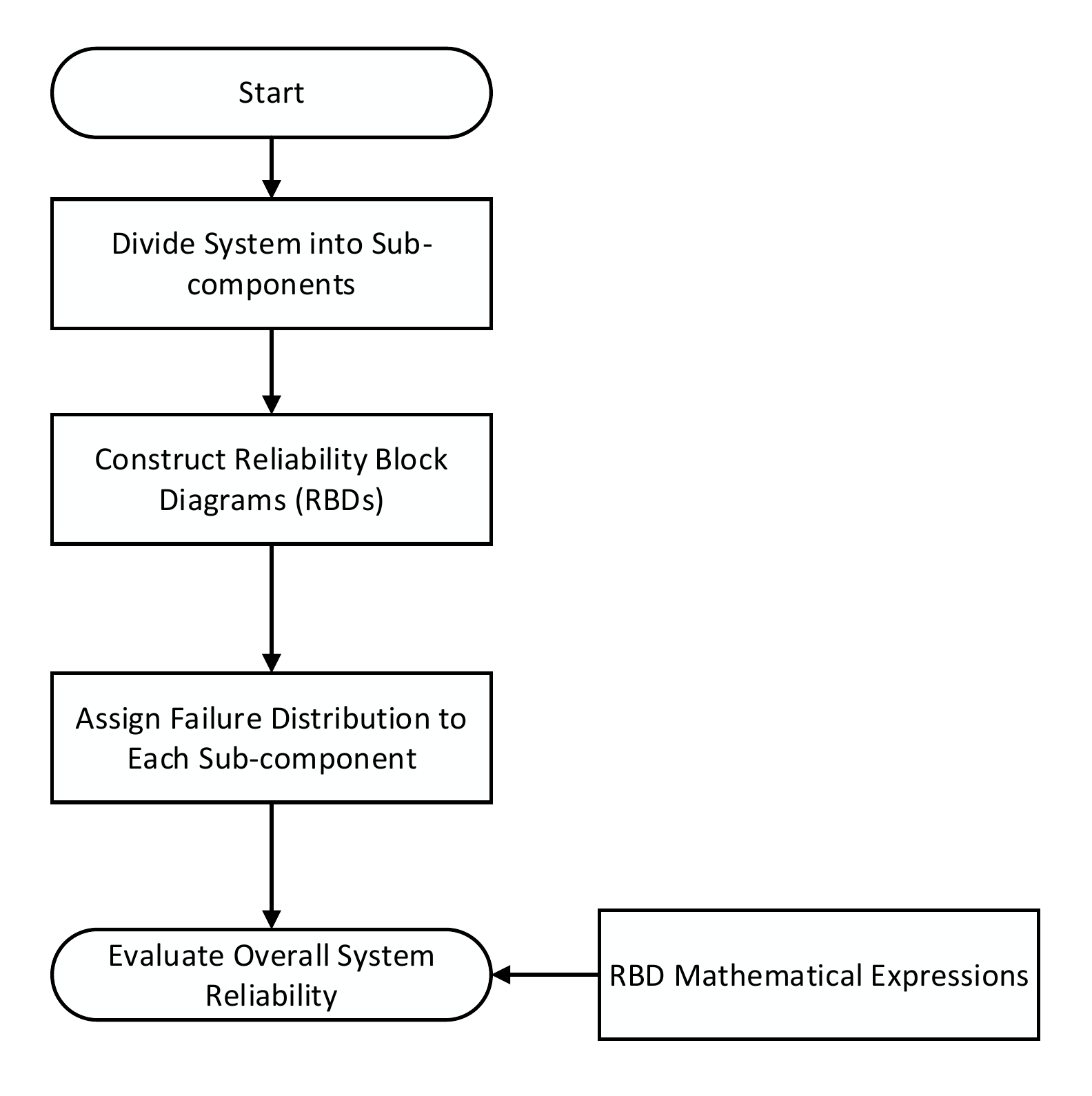}\\
  \caption{Reliability Block Diagram Analysis Process}\label{RBD_pro}
\end{figure}

The analytical RBD-based reliability analysis process is illustrated in Figure \ref{RBD_pro} and it begins by the logical partitioning of the system into its components. This partitioning may be based on functional behavior or actual connections of the components in the system. This is followed by the construction of an RBD and the assignment of failure distribution to each individual component. Usually, the \textit{Exponential} or \textit{Weibull} distributions, with failure rate $\lambda$ and time-to-failure random variable, say $X$, are used in order to express the reliability or availability of these individual components. The dependability of each component is then used in order to determine the reliability or availability of the overall system by utilizing the mathematical expressions that are presented in Equations 1-5.

\subsubsection{Fault Tree Analysis}
\label{sec:FTA}

\begin{figure}[!h]
  \centering
  \includegraphics[width=0.7\textwidth]{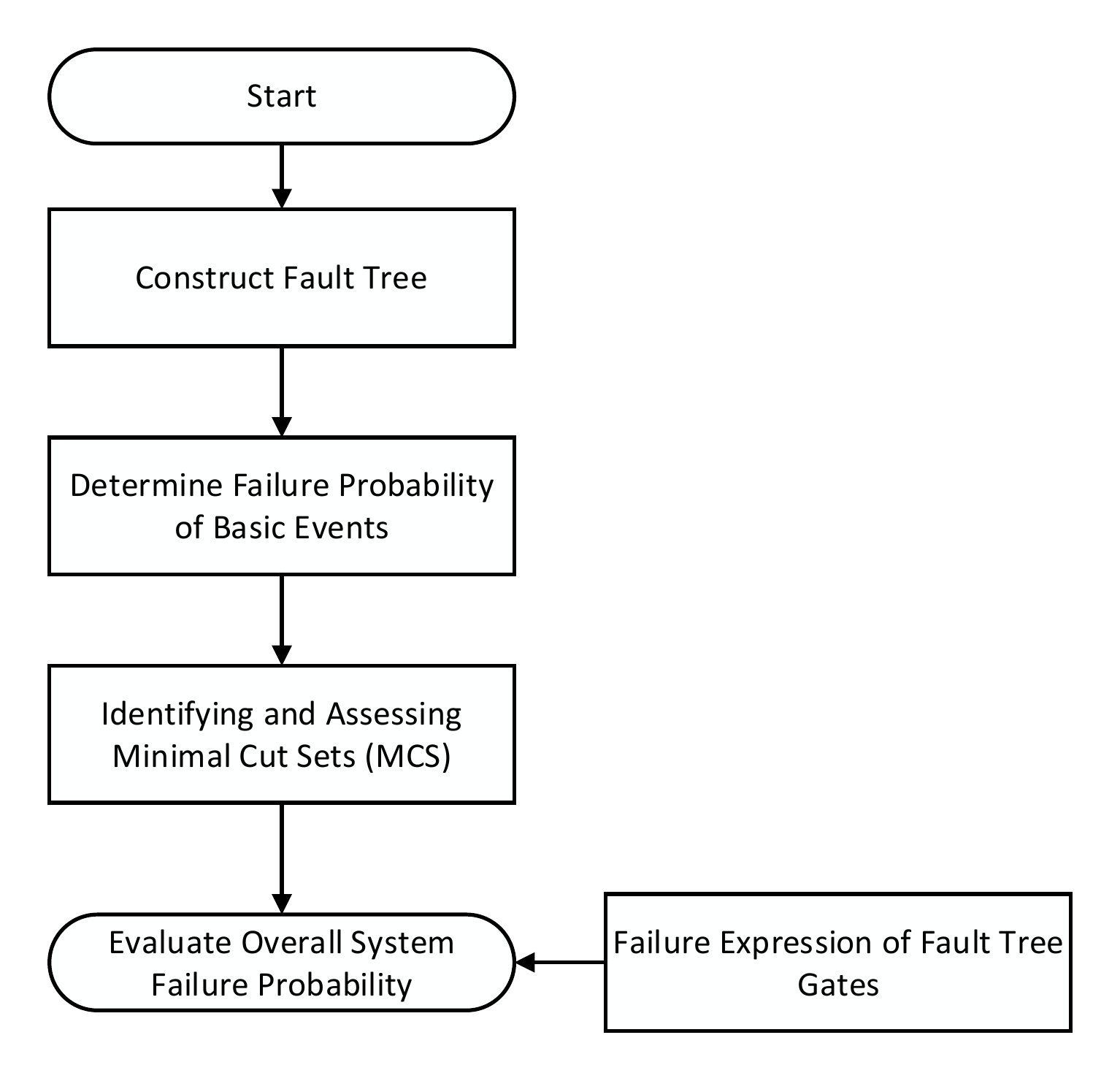}\\
  \caption{Procedure of Conducting Fault Tree Analysis}\label{FTA_pro}
\end{figure}

In Fault Tree analysis (FTA), each FT gate has an associated failure probability expression as shown in Table \ref{FT_table}. These expressions can be utilized to evaluate the reliability of the system. The first step in the FTA, as illustrated in Figure \ref{FTA_pro}, is the construction of the FT of the given system. This is followed by the assignment of the failure distributions to basic $cause$-$events$ and the identification of the Minimal Cut Set (MCS) failure events, which contribute in the occurrence of the top event. These MCS failure events are generally modeled in terms of the \textit{Exponential} or \textit{Weibull} random variables and the Probabilistic Inclusion-Exclusion (PIE) principle \cite{trivedi2008probability} is then used to evaluate the probability of failure of the given system.

Mathematically, the PIE can be expressed as follows:
\begin{equation}
P(\bigcup_{i=1}^{N}A_{i}) = \Sigma_{k=1}^{N}((-1)^{k-1}\Sigma_{I\subset\{1,...,n\} \\ \& |I|=k}P(\bigcap_{i\in I})A_{i})
\end{equation}

\noindent where $A_{i}$ corresponds to the $i^{th}$ cut set event.

\begin{table}[!ht]
\centering
\caption{Probability of Failure of Fault Tree Gates}
\begin{tabular}{|l|l|}
\hline
\emph{Fault Tree Gates} & \emph{Failure Probability Expressions} \\
\hline
\hline
\parbox[c]{1em}{
\includegraphics[width=0.8in]{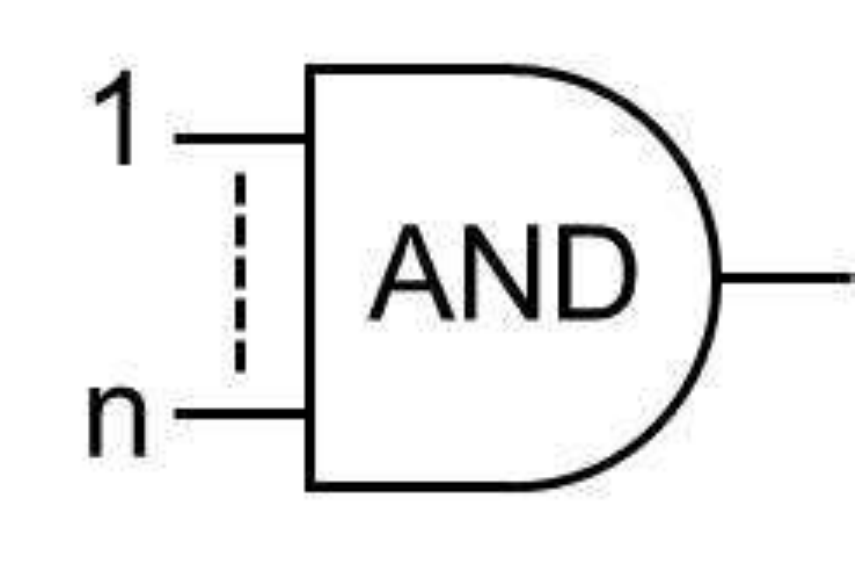}}& $\!\begin{aligned}[t]
 F_{AND\_gate}(t)
    & = Pr (\bigcap_{i=2}^{N}A_{i}(t))
    = \prod_{i=2}^{N}F_{i}(t)
    \end{aligned}$
\\
\parbox[c]{1em}{
\includegraphics[width=0.8in]{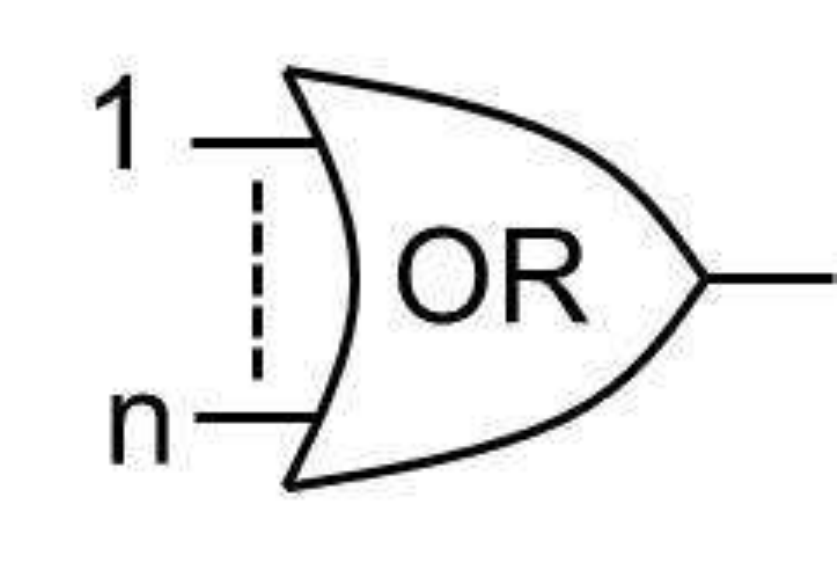}}& $\!\begin{aligned}[t]
 F_{OR\_gate}(t)
    & = Pr (\bigcup_{i=2}^{N}A_{i}(t))
    = 1 - \prod_{i=2}^{N}(1 - F_{i}(t))
    \end{aligned}$
\\
\parbox[c]{1em}{
\includegraphics[width=0.8in]{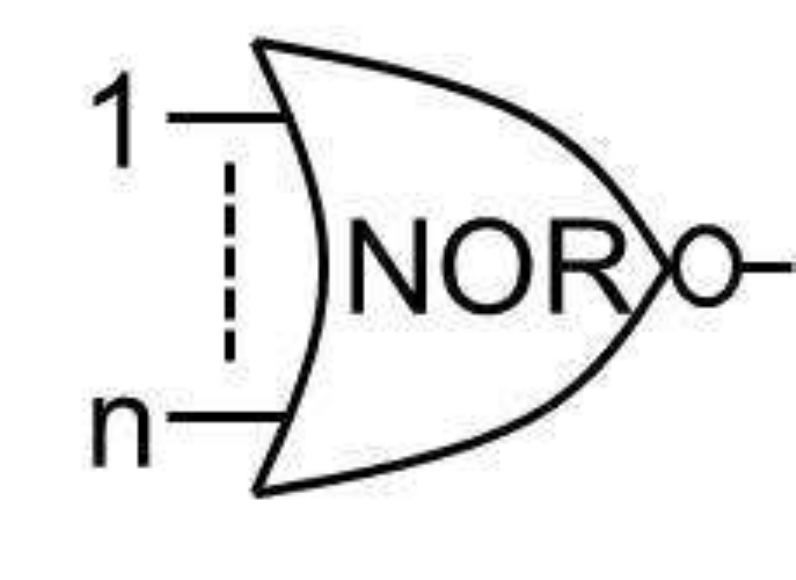}}& $\!\begin{aligned}[t]
 F_{NOR}(t)
    & = 1 - F_{OR}(t)
    = \prod_{i=2}^{N}(1 - F_{i}(t))
    \end{aligned}$
\\
\parbox[l]{1em}{
\includegraphics[width=0.8in]{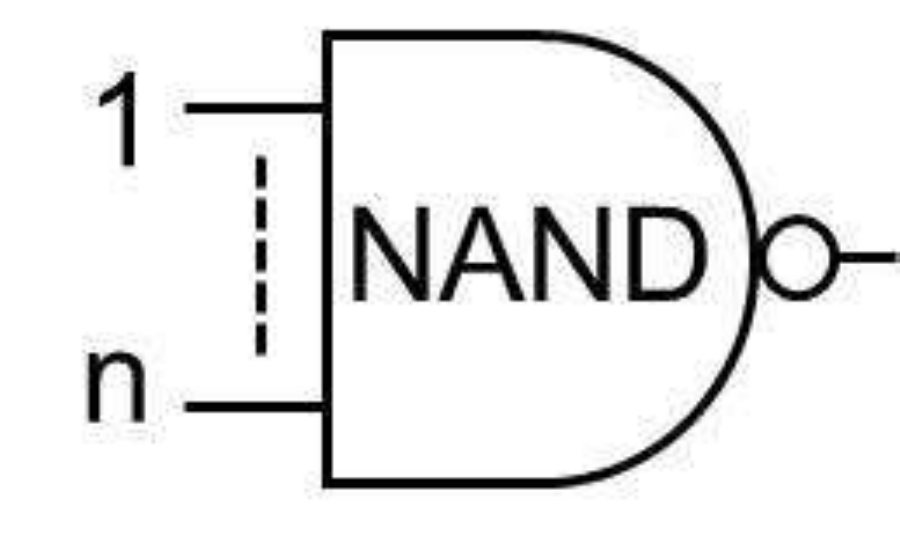}}&  $\!\begin{aligned}[t]
F_{NAND}(t) & =  Pr (\bigcap_{i=2}^{k}\overline A_{i}(t) \cap \bigcap_{j=k}^{N}A_{i}(t)) \\ 
&= \prod_{i=2}^{k}(1 - F_{i}(t)) *\prod_{j=k}^{N}(F_{j}(t))\end{aligned}$   \\
\parbox[c]{1em}{
\includegraphics[width=0.8in]{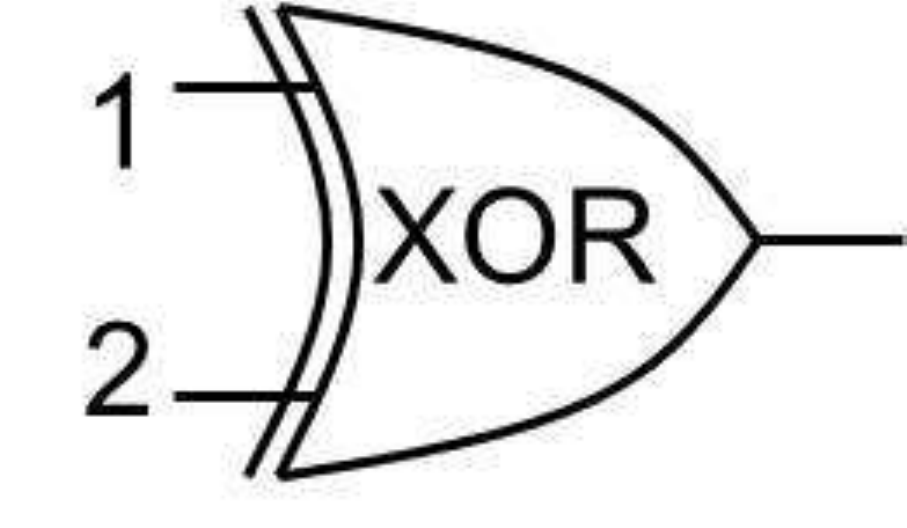}}& $\!\begin{aligned}[t]
 F_{XOR}(t)&= Pr(\bar{A}(t)B(t) \cup A(t)\bar{B}(t)) \\ 
 &= F_{A}(t)(1 - F_{B}(t)) + F_{B}(t)(1 - F_{A}(t))\end{aligned}$    \\

\parbox[l]{1em}{
\includegraphics[width=0.8in]{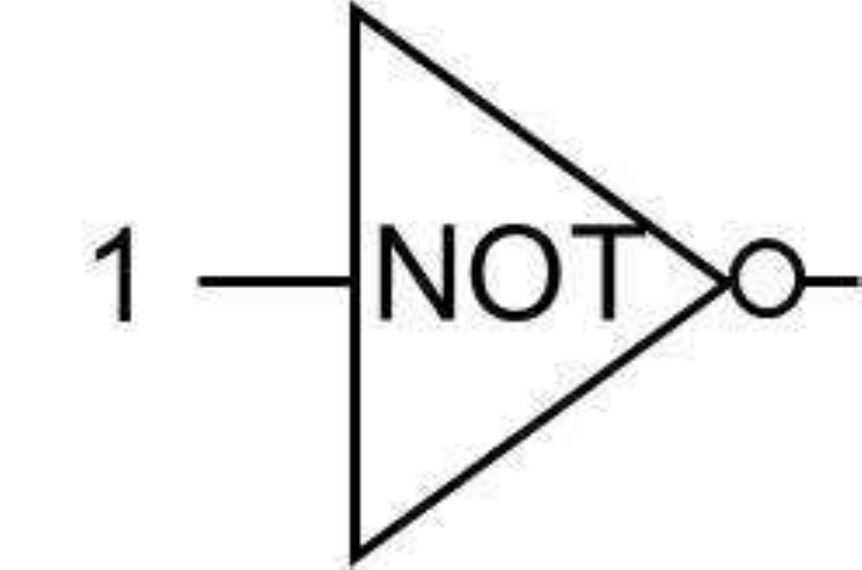}}& $\!\begin{aligned}[t]
 F_{NOT}(t)&= Pr(\overline{A}(t)) =(1 - F_{A}(t))\end{aligned}$     \\
\hline
\end{tabular}\label{FT_table}
\end{table}

\subsubsection{Markov Analysis}
\label{sec:MA}

It is a stochastic process $X(t)$ which is defined over the discrete state-space $\omega$.  A process $X(t)$ can be considered as a Markov process only if, given the sequence of time instants $(0<t1<t2...t_{m})$, the probability of a system being at state $x^{(m)}$ at time instant $t_{m}$ is only dependent upon the previously occupied state $x^{(m-1)}$ at time instant $t_{m-1}$. This property provides the basis for Markov process and is known as the Markov property. Mathematically, it can be expressed as:

\begin{equation}
\centering
\begin{split}
& Pr(X(t_{m}) =x^{(m)}|X(t_{m-1}) =x^{(m-1)},...,X(t_{1}) =x^{(1)}) \\ = &Pr(X(t_{m}) =x^{(m)}|X(t_{m-1}) =x^{(m-1)})
\end{split}
\end{equation}

\begin{figure}[!h]
  \centering
  \includegraphics[width=0.7\textwidth]{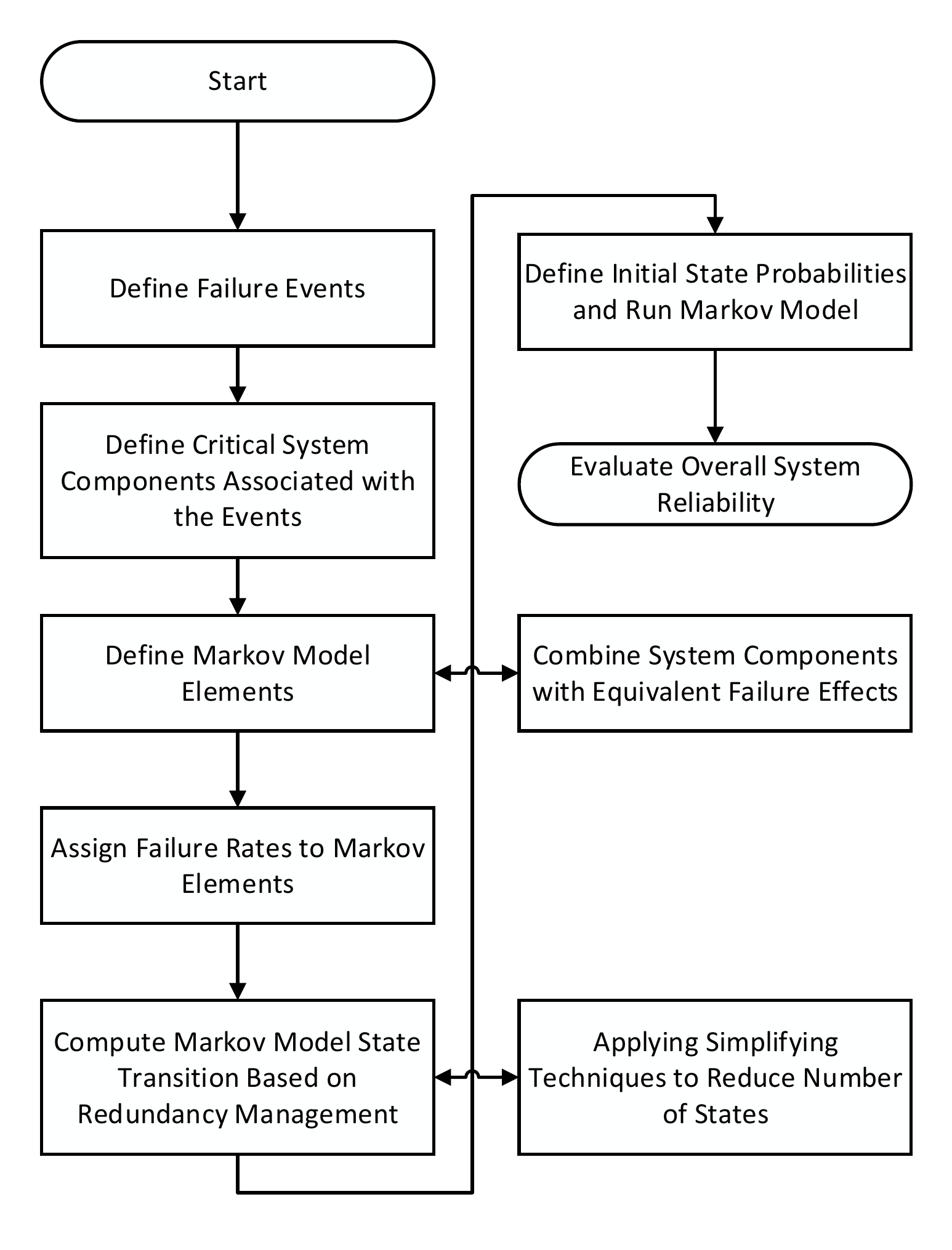}\\
  \caption{Markov Chain Analysis Design Process}\label{Markov_pro}
\end{figure}

\noindent For DTMC, these time instants are discrete while in CTMC they are continuous.

The main steps involved in the Markov chain analysis process are shown in Figure \ref{Markov_pro}. The process starts from defining the failure events and the identification of critical system components of the system. This is followed by first defining the basic Markov model elements of the critical components of the system and then assigning the failure probabilities to these basic elements. It is usual that the states in the Markov model are very large so the next steps involves the reduction of state-space of the Markov model by using a theorem \cite{john1967finite}, which states that given the steady-state probabilities of original Markov chain, the probabilities for smaller chain are proportional to the corresponding states in the original Markov chain. When the modeling is complete, the analysis starts by defining the initial probabilities and then running the system to evaluate the overall system reliability or availability.

\subsubsection{Fuzzy Logic}
\label{sec:FL}

The system reliability analysis often requires the use of uncertain data, subjective judgment and the approximate system models. Fuzzy logic \cite{klir1995fuzzy} provides an effective tool for reliability analysis when the analysis does not require precision or accurate results. Fuzzy logic can be considered as the generalization of classical logic and is used to model those systems in which the data used is not precise or the inference rules are formulated in a very general way \cite{bowles1995application}. In classical propositional logic \cite{segerberg1982classical}, there are only two values  \textit{0} or \textit{1}, while the fuzzy logic provides the whole continuum of truth values for propositional logic.

A fuzzy logic is characterized by a set known as $Fuzzy Set$ and its associated member function. A membership function is used to assign degree of membership to each member $x \in X$. Consider a membership function $\lambda(x)$ on interval [0,1], where 0 represents the absence of membership of element $x$ in the fuzzy set $A$ and 1 represents the complete membership in set $A$. While, the values between 0 and 1 show the partial membership of the element $x$ in the fuzzy set $A$. Mathematically, it can be written as:

\begin{equation}\label{eq:fuzzy_set}
\centering
 A = \{(x,\lambda(x))| x \in X \wedge 0 \leq \lambda(x) \leq 1\}
\end{equation}

In reliability analysis, the evaluation of probability of a system either working correctly or in the state of failure is our prime concern. However, in reality, their may be a situation in which system performance degrades before it enters in a failure state and in between there are ranges of states in which the system is partially functional. For the aforementioned situation, the fuzzy logic has been found to be in a better position to define the reliability events for a given system. Fuzzy logic can provide reliability analysis by utilizing RBD \cite{bowles1995application} or FT \cite{tanaka1983fault} models.

\subsubsection{Bayesian Network Analysis}
\label{sec:BNA}
In a Bayesian Network (BN), the dependency of a component can be determined by the link between two nodes in a child parent relationship. The child represents the dependent component of the parent node. Therefore, the success of the child node is conditional on the success probability of the parent node. The Bayes theorem \cite{vapnik1998statistical} is used to evaluate the conditional probabilities of the child node by considering the probability value associated with the parent node. Also, if the link between any two nodes is absent then it represents that these components do not interact for system failure and can be considered as independent. So the probabilities of these components should be evaluated separately.

Consider a BN over $U = X_{1},X_{2},...,X_{n}$ where $X_{1},X_{2},...,X_{n}$ are the nodes. Then, based on the chain rule, the joint probability $Pr\{X_{1},X_{2},...,X_{n}\}$ is given as follows \cite{cai2012using}:
\begin{equation}\label{eq:bn}
Pr(U) = Pr\{X_{1},X_{2},...,X_{n}\} = \prod_{i=1}^{n} Pr(X_{i}|\pi_{i})
\end{equation}

\noindent where $\pi_{i}$ is the set of parents of node $X_{i}$. The probability distribution of a random variable at a particular node can be found by marginalizing the joint probability distribution, which is presented in Equation \ref{eq:bn}. This process is known as marginalization and it can be used to compute the failure probability and reliability of the given systems \cite{cai2012using}.

\subsection{Simulation Tools}
\label{sec:simul_tools}
 Numerical methods based simulation tools have been extensively utilized for the reliability analysis of communication networks. The main idea behind these tools is to model the random phenomena by using pseudo-random number generation methods and thus use them to depict probability distributions of failures in reliability analysis. Thereafter, numerical analysis techniques can be used to judge the overall reliability of the given model, i.e., RBD, FT or MC, using its corresponding mathematical relationship, verified using the analytical methods described above. There are many simulation based reliability analysis tools, which are commercially available and provide many attractive features including graphical editors in which complex systems can be modeled with great ease. These tools take the system reliability model and associated system components failure distributions as an input from the user and return the overall reliability or failure probability of the system numerically. The details of some of these tools are as follows:

 \subsubsection{SHARPE}

Symbolic Hierarchical Automated, Reliability and Performance Evaluator (\textit{SHARPE}) \cite{SHARPE_14} is a toolkit which can be used to analyze reliability, availability and performability of systems. It supports combinatorial reliability techniques, such as FTs and RBD, as well as state-based modeling techniques, such as Markov chain. \textit{SHARPE} also allows users to choose among different reliability analysis techniques and can provide results in the form of a distribution function or as a mean and probability.

\subsubsection{Isograph}
\textit{Isograph} \cite{Isograph_14} provides a package of software, such as \textit{Availability Workbench, HAZOP+, NAP} and \textit{AttackTree+}, which are  capable of providing analysis for reliability, maintainability, availability and safety \cite{Isograph_14}. They utilize the reliability modeling techniques, such as FT and RBD, to predict the reliability, throughput and life cycle cost of the system. \textit{Isograph} allows modeling the system concisely in an easy to understand graphical format.

\subsubsection{RAPTOR}
 In this tool, the graphical user interface allows the user to draw the RBDs and describe the components failure or repair behavior \cite{raptor_14}. The user can place the components and make the connection between these components using the links and the nodes. Any combination of series-parallel RBD structures can be designed using the powerful workspace provided by \textit{RAPTOR}. The failure behavior of the system or components can be described by using 16 built-in distributions. \textit{RAPTOR} can provide reliability prediction and availability and maintainability analysis. In addition, it supports life cost analysis and sensitivity analysis as well.

\subsubsection{ReliaSoft}
Like \textit{Isograph}, \textit{ReliaSoft} also provides a wide variety of software, such as \textit{WEIBULL++, ALTA, DOE++} and \textit{BLOCKSIM} \cite{Reliasoft_14}. The main features provided by these software tools are system reliability analysis, identification of critical components, optimum reliability allocation, system availability analysis and throughput calculation. In addition, these tools provide a user-friendly interface to utilize complex and powerful mathematical models for quantitative accelerated life testing data analysis.

\subsubsection{Galileo}
\textit{Galileo} \cite{sullivan1999galileo} is a dynamic fault tree modeling and analysis tool that combines the binary decision diagram (BDD) and Markov methods with rich user interfaces. Some distinguishing features of the tool are: (i) Users can edit a fault tree in either a textual or graphical representation; (ii) It provides very usefull editing features,  such as zoom, find-and-replace,
print preview, cut-and-paste; and (ii) It comes with a very comprehensive  on-line documentation through an embedded
internet browser and World-Wide Web pages.

A part from above mentioned tools, some XML-based specification, like OPC unified architecture \cite{cavalieri2013analysis}, has been developed to realise the communication among industrial applications. The main feature is to analyze the overhead of client/server data exchanges.

\subsection{Formal Methods}

Formal methods \cite{clarke1996formal} are computer-based mathematical analysis techniques. The main idea is to represent the system's behavior and its desired properties in appropriate logical frameworks and develop a relationship between these two by mathematical reasoning or rigorous state exploration techniques in a computer. This rigorous analysis approach of these techniques allow us to verify the system's properties in a more thorough fashion than that of empirical testing. In this section, we describe a brief overview of some formal methods that have been utilized extensively for the reliability analysis of communication networks.

\subsubsection{Model Checking}
Model checking \cite{clarke1999model} is one of the most commonly used formal verification techniques for the verification of finite-state concurrent systems, such as sequential circuit designs and communication protocols. In this technique, a given system is first represented as a finite state automata $M$ and then the verification is done by verifying temporal logic formulas $\phi$ exhaustively over the complete state-space of $M$ \cite{clarke1999model}.

\begin{equation}\label{model_checking}
  M \models \phi
\end{equation}

Model checking \cite{clarke1999model} has been increasingly utilized for analyzing safety and mission critical applications and it has gained significant popularity in the industry because the verification process can be fully automated and counterexamples are automatically generated if the property being verified does not hold \cite{ouimet2007formal}. However, the state-space explosion problem  \cite{ouimet2007formal}, i.e., the problem of dealing with the computational overhead of exhaustive state-exploration of complex systems, is one of the inherent limitations of this technique.

\textit{PRISM} \cite{kwiatkowska2011prism} is an open-source probabilistic model checker, which has been extensively used for the reliability analysis of a variety of communication systems. It provides support for building and analyzing several types of probabilistic models, such as discrete and continuous-time Markov chains, Markov decision processes, and extensions of these models with rewards. \textit{PRISM} constructs the corresponding probabilistic model of the given system in the form of states and transitions. For reliability analysis, the states represent the functioning or failure of the system or its components and transitions represent a possible evolution of the system from one configuration or failure to another over time. Transitions are labeled with quantitative information, such as failure rates of exponential distributions. Once this probabilistic model is constructed, it can be used to analyze a wide range of quantitative properties, related to reliability, of the original system.

\subsubsection{Petri Nets}
Petri Net (PN) \cite{peterson1981Petri} is a bipartite directed graph consisting of disjoint sets of places $P$ and transitions $T$. The former, which is represented by circles, models the condition while the latter, signified by bars, represents the events or activities that may occur in the system. The directed arcs $(P X T)$ and  $(T X P)$, represented by arrows, describe the input places $P$ for the transitions $T$ and output places $P$ for the transitions $T$, respectively. Places may be empty or contain more than one token that is drawn by a block dot and term $marking$ represents the tokens over the set of places.

A transition is said to be enabled, in a given marking, if all its input places contain at least one token. An enabled transition can be $fired$ and as a result a token will be removed from the input places of the transition and added to its output places.

A stochastic variant of a Petri Net is commonly used for the reliability analysis \cite{liu1997application}. A Stochastic Petri Net (SPN), based on CTMC, is a timed PN where each transition is characterized by an \textit{exponential} random variable. A reward function can be used to define the reliability metrics of an SPN based model as follows:
\begin{equation}\label{eq:SPN}
\centering
r_{m} =
\begin {cases}
1 \hspace{4pt} \text{if} \hspace{4pt} M \in O \\
0 \hspace{4pt} \text{if} \hspace{4pt} M \in F
\end{cases}
\end{equation}

\noindent where $r_{m}$ is a state reward, which splits the set of reachable markings of an SPN into two subsets: $O$ represents the operational state of the system and $F$ represents the failure state. The instantaneous steady state and the interval availability metrics can be calculated using this state reward function.

A stochastic activity networks (SAN) \cite{sanders2001stochastic,cavalieriquantitative}, i.e., a variant of SPN, are specifically being developed for performance, dependability, and performability evaluation of a wide range of real-world systems. For instance, an integrated modeling approach for fault tolerant systems
based on SAN has been used for the evaluation of system’s performance parameters, such as reliability, availability and maintenance costs \cite{maza2014stochastic}. Similarly, the Hierarchical Stochastic Activity Networks (HASN), an extension of SAN, has been used to quantitatively analyze the performance of QoS for intensive video traffic over the 802.11e standard used in WLANs \cite{perez2015evaluation}.  SANs have also been used to analyse the performance of the KNXnet/IP congestion control mechanism \cite{cavalieri2014limiting}.

In Colored Petri Nets (CPN) \cite{jensen1983high}, labels are used to distinguish tokens and it allows to implement different firing policies for different tokens. Moreover, a CPN can provide a dynamic change of the token labeling by specifying different labels to the tokens when they enter and leave the transitions. The ability of a CPN to move a token through a transition among the places while altering the labels during this process provides a flexible and powerful mechanism of tracking relevant changes in the system \cite{volovoi2004modeling}. This mechanism allows the PN to model commonly captured behaviors of RBD and FT  \cite{codetta2005conversion}.

\subsubsection{Higher-order-Logic Theorem Proving}
 Higher-order Logic (HOL) is a system of deduction with a precise semantics and is expressive enough to model any system that can be expressed in a closed mathematical form. Moreover, in conjunction with general purpose proof assistant, such as \textit{Coq} \cite{bertot2013interactive}, \textit{PVS} \cite{owre1992pvs} and \textit{HOL4} \cite{slind2008brief}, HOL can be utilized to develop automatic and interactive proof environments for reasoning about system properties. The formal analysis carried out in this manner is logically sound and provides an irrefutable proof that a system satisfies its requirements.

 A number of higher-order-logic formalizations of probability theory are available in higher-order logic (e.g., \cite{hurd_02,hasan_phd_08, mhamdi2012information}) and have been utilized to
verify sampling algorithms of a number of commonly used discrete \cite{hurd_02} and continuous random variables \cite{hasan_cade_07} based on their probabilistic and statistical properties
\cite{hasan_icnaam_07,hasan_fm_09}. Moreover, this formalization has been used to conduct the reliability analysis of a number of applications, such as memory arrays \cite{hasan_tc_10}, soft errors \cite{abbasi_13b}, electronic components \cite{abbasi_13} and oil and gas pipelines \cite{WAhmad_CICM14}.

\section{Reliability Analysis of Communication Networks with RBD Models}
\label{sec:AnalysisWithRBD}

As described earlier in Section \ref{sec:depend_model}, RBD provides a graphical model of a system by representing its components as blocks and their interrelationship by using the directed lines between the blocks. This obtained graphical model can then be utilized to evaluate overall system reliability by using its individual component reliability. RBD models have been extensively utilized, due to their ability to represent complex systems in a simplified graphical form, for the reliability analysis of communication networks, including wireless sensor networks, ad-hoc networks, cellular networks, multiple computer networks, cloud computing and power grid communication networks. This section primarily presents a survey on the RBD-based reliability analysis in the domain of communication networks. The information is classified on the basis of reliability analysis techniques, i.e., analytical, simulation tools and formal methods.

\subsection{Analytical Methods with RBD}

\textit{Fault tolerance} \cite{fault_tolerance_45203} has been an extensively utilized approach in communication networks. The main idea is to mask the expected network faults by using additional standby network components. Similarly, the nodes in the communication networks, such as wireless sensor networks (WSNs), multistage interconnection networks (MINs), optical network and bus architecture, are made redundant by connecting extra nodes in series or parallel to the underlying network, thus making these networks tolerant to faults. Consequently, the reliability of communication networks is significantly increased because a standby node can replace any node that stops working. RBD-based models are extensively utilized to carry out the reliability analysis of these fault tolerant communication networks. For example, the fault  tolerance approach has been used on the wireless sensor nodes of a WSN \cite{bein2005reliability}  with the objective of making the network more reliable. To observe the effects of fault tolerance on network reliability, the sensor nodes are modeled with a series-parallel RBD configuration in this case.
Similarly RBDs have been used to assess the reliability of many other fault tolerant communication networks including wireless sensor networks (WSNs)  \cite{ghaffari2008fault},  multistage interconnection networks (MINs) \cite{bistouni2014pars}, wide area networks (WANs) \cite{jiajia}, numerical protection systems \cite{yang2009impact} and electric power communication networks \cite{chen2009design}.

One of the essential performance parameters of any communication network is its ability to reliably transmit the communication data. Therefore, many techniques have been developed to improve the reliability of the \textit{data transmission} in communication networks at the software as well as hardware levels. Some of the commonly used methodologies include applying filters on the data to mitigate the irrelevant data \cite{filtering}, data transmission strategies \cite{data_transmission_strategies} \cite{1_data_transmission_strategies}, making the systems fault tolerant \cite{fault_tolerance} and introducing feedback systems \cite{feedback}. For reliability analysis of these kinds of enhanced communication network systems, RBDs have been found to be more preferable than other modeling techniques due to their ability to effectively capture the effect of failures of individual network components upon the overall network. In WSNs, data transmission mechanism is improved by applying various operations on the data which include data retransmission, filtering out the irrelevant data, removing extra information and checking the vulnerability of the data \cite{shaikhmodeling}. Since, these operations are applied on the data one-by-one so RBD can model these operations using a combination of series and parallel RBD configurations. Similarly, in the case of multistage interconnection networks (MINs), the data transmission system is made reliable by using switching stages, which have been modeled by using series-parallel and parallel-series RBD configurations \cite{bistouni2014analyzing}.

RBDs have also been used to predict the level of reliability required to mitigate the performance degradation effects, including unreliable data transmission, service unavailability, \textit{components dependencies} and inappropriate data throughput, in many communication networks \cite{nojo1993incorporating}, such as network service systems \cite{xin2013network} and smart grid stations \cite{wafler2013combined}. For instance in smart grid substations, the information and communication technology (ICT) infrastructure is usually comprised of components,  such as meters and antennas, which are dependant on each other and thus their reliability dependancy has been captured using the series-parallel RBD configuration to analyze the reliability and availability of the ICT infrastructure \cite{wafler2013combined}.

The reliability of \textit{communication protocols}, which are primarily a set of rules that make communication effective and reliable between nodes or systems, has also been extensively studied using RBDs \cite{Argyriou200660} \cite{xiangning2007improvement} \cite{kramer2002ipact}. Similarly, RBDs have also been used to evaluate the reliability of newly designed protocols, such as reliable energy-efficient cluster-based protocol (REECP) and routing protocol for WSNs \cite{zhou2007reecp} \cite{damaso2014reliability}. For instance, the reliability of the cluster-based communication protocol REECP, which effectively reduces the energy consumption of a WSN, is evaluated by modeling the cluster heads and sensor nodes with series, parallel and series parallel RBD configurations \cite{zhou2007reecp}.

%email
\textit{Network topologies} are the schematic arrangements of the network components to establish an efficient communication system. Network topologies have an immense effect on the reliability of communication systems and thus it is required to adopt the most reliable network topology for a particular network \cite{topology}. To observe the effects of a network topology on the network reliability, RBD modeling techniques are very commonly used. The topology is transformed into a RBD model by modeling the network nodes with the most suitable configuration. For instance, in multi-computer systems, a hypercube topology of the processing elements (PEs) is established to make the interconnection network (IN) fault tolerant. The hypercube computer network is then modeled with RBD configurations and analyzed analytically for the reliability evaluation of multi-computer system  \cite{abd2014reliability}.

Ethernet architecture based communication networks have been developed to introduce interoperability among communication devices supplied by different vendors in substation automation systems (SAS) \cite{kanabar2009reliability}. The reliability and availability of these ethernet architectures is assessed using the series-parallel RBD configuration \cite{kanabar2009reliability} in order to analyze the overall reliability and performance of the communication system of SAS.

%email
Fuzzy logic \cite{bowles1995application} and Bayesian networks \cite{langseth2007bayesian} can further extend the analysis capabilities of RBD models and thus have been used in conducting reliability analysis of communication networks. The Bayesian network based RBD technique can model the \textit{dynamic nature} of the systems and the fuzzy RBD hybrid technique can model the \textit{uncertainties} of the system. Bayesian network based RBD approach is also utilized to evaluate the reliability of complex communication networks by centering the complex parts of the system by its unique reverse reasoning technique. Moreover, Bayesian networks also facilitate to capture the probabilistic interaction of the network components. For instance, in wide area protection system \cite{hai2012analysis}, the reliability of the communication system depends on the subsystems that are system protection centered, backbone networks and regional communication systems. These subsystems are first modeled with RBDs and then their RBD models are transformed into their corresponding Bayesian networks for reliability analysis. The polytree system has been analyzed on the same lines in  \cite{torres1998bayesian}. Fuzzy logic based RBD approach can effectively model the complex systems that pose linguistic uncertainties. It can be utilized to evaluate the reliability as well as the lifetime of the systems. For instance in WSNs, the behavior of the lifetime of a single sensor node follows normal Gaussian distribution and hence it is modeled as a Gaussian membership function with uncertain standard deviation for the reliability and lifetime analysis of WSN \cite{shu2008wireless}.

\subsection{Simulation Based Analysis with RBD}
Simulation tools, such as \textit{Galileo} \cite{coppit2000galileo}, \textit{HARP} \cite{DarrellW.:2003:HHA:886548} and \textit{SHARPE} \cite{Sahner:1986:RMU:868535}, have been found to be very effective for the reliability analysis of large communication systems with many nodes.

Communication networks, such as vehicular ad-hoc networks  \cite{dharmaraja2012reliability}, ICT infrastructures \cite{malek2008analytical} and VoIP based communication systems \cite{palade2010reliability}, are very complex as they are composed of many sub-systems, such as hardware systems, software systems, virtualized systems, communication mediums etc. Failure in any system may lead the whole network to move to some undesirable state. Reliability analysis of these systems to ascertain the effect of a sub-system or a component failure to the overall system is a matter of prime concern. RBDs are considered to be quite preferable to model such kind of complex networks for their reliability evaluation and the analysis can be made easy by using simulation tools. For instance, the failure behavior of hardware components of on board unit (OBU) and road side unit (RSU) of VANETs \cite{dharmaraja2012reliability} have been used with series and series-parallel RBD configurations in the SHARPE software tool to evaluate the reliability of the overall system. Similarly, for large communication networks, such as SCADA system of a power grid station,  which consist of many nodes and links, the RBD based reliability analysis is conducted using the ITEM simulation tool  \cite{bobbio2010unavailability}.

Many communication networks \cite{pokorni2011reliability} \cite{held2003consideration} are error prone and exhibit some \textit{repairing strategies} either on component level or on system level. Networks with repairing strategies are complex and thus their reliability assessment becomes quite challenging. Since, RBD can model the fundamental components of any system with their failure and repair rates, it is considered to be a more preferable approach to model the networks with repairing strategies. For instance in a long haul network \cite{held2003consideration}, the three repairing/protection strategies, which include span protection, path protection and protection cycle, are modeled with RBD and reliability analysis is conducted using Monte Carlo simulation.

\subsection{Formal Methods with RBD}

The higher-order logic theorem prover \textit{HOL4} has been used for RBD analysis and some preliminary results related to the reliability analysis of oil and gas pipelines, composed of serially connected sub-components, are reported in \cite{WAhmad_CICM14}. In particular, this work utilizes the probability theory \cite{mhamdi_lebsegue}, available in the \textit{HOL4} theorem prover, to formalize reliability, the exponential random variable and the series RBD. These foundations are then used to formally verify a generic expression of a simple pipeline structure that can be modeled as a series RBD with an exponential failure time for individual segments. Recently, this work is extended for the formalization of commonly used RBDs, which is then utilized to conduct formal reliability analysis of a complex virtual data center \cite{ahmed2016formalization}. The HOL formalization of RBDs has also been used for availability analysis of satellite solar arrays \cite{ahmed2016formal}. To our best knowledge, none of the work present in the current literature utilizes theorem proving for the reliability analysis of communication networks.

Petri Nets and its variants are widely used as a reliability analysis tool in the domain of communication networks due to their ability to efficiently handle large problems of dynamic nature.  For instance, the live migration process in cloud computing networks makes the system dynamic and thus yields to a complex RBD model, which can be effectively handled using Petri Nets with the support of commercial tools, such as \textit{SNOOPY} \cite{snoopy} and \textit{CPN} \cite{beaudouin2001cpn}. Given the dynamic nature of visualization, due to the presence of hardware systems, software systems, live migration techniques, resource allocation algorithms and concurrent failures, virtualized networks are frequently modeled with RBDs, which are then transformed to Petri Nets for the reliability analysis \cite{lira2013virtual} \cite{fernandes2012dependability} \cite{wei2011dependability}. The reliability of communication networks with \textit{redundancy mechanisms} has also been efficiently analysed using RBD based Petri Nets \cite{guimaraes2011dependability} \cite{6014733}. Similarly, given the dynamic nature of Service of Supervision, Control and Data Acquisition System (SCADA) communication of power grid stations, their availability is also analysed using Petri Nets  \cite{bobbio2010unavailability}. \textit{Power management polices}, such as sleep/wake policies, greatly reduce the power consumption of the overall communication network but the behavior of the network becomes dynamic and thus Petri Nets can be used for the reliability analysis of such systems. For example,  the reliability analysis of WSNs that work on sleep/wake policies has been done by capturing the dynamic behavior of the sensor nodes by a DRBD model, which is transformed into Petri Nets for easy analysis. Petri Nets have also been used to ensure the security/safety aspects of networks in terms of reliability and availability by analyzing the safety/security aspects of network protocols,  such as internet voting systems \cite{omidi2012modeling} and high-speed trains \cite{lijie2012verification}. Petri Nets are also used quite frequently to estimate the capacity of the communication networks required to meet the needs of the customers. For this purpose, the Petri Nets based reliability analysis of the designed network architecture is performed \cite{Lima201427} \cite{queiroz2013performability}.

A summary of the literature on reliability assessment of communication networks using RBD-based modeling with analysis performed through analytical and simulation methods for RBD models is presented in Table \ref{RBD_Ana_Sim} whereas the literature involving reliability analysis using formal methods is summarized in Table \ref{RBD_formal}.

\begin{longtable}{p{0.25\linewidth}p{0.06\linewidth}p{0.6\linewidth}}
\caption{Representative summary of RBD-based reliability assessments using various \underline{\textit{Analytical/Simulation}} analysis techniques}\label{RBD_Ana_Sim}\\
%\begin{tabular}{p{3cm}p{1cm}p{10cm}}

\toprule

 \textbf{Network Characteristic}   & \textbf{Ref.}  & \textbf{Brief summary} \\
\midrule
\small
Fault Tolerance     & \cite{bein2005reliability} & A fault tolerant WSN system is modeled using the series-parallel configuration of RBD\\
                    & \cite{ghaffari2008fault}  & A framework is presented to incorporate the data transmission problem and is modeled using the series configuration of  RBD\\
                    & \cite{bistouni2014pars} & The pars network, i.e., a fault tolerant interconnection network, is presented and its reliability analysis is performed using RBD\\
                    &  \cite{jiajia} & The reliability analysis of different protection systems, for the PON architecture,  is performed using RBD technique\\
                    & \cite{yang2009impact} & The reliability analysis of smart grid substation is performed using the RBD technique while considering the dynamic behavior of the systems\\
                    & \cite{chen2009design} & The reliability analysis of different bus architectures, which are deployed in substation communication system, is performed using the RBD technique\\
\hline

Data Transmission System & \cite{shaikhmodeling} & The data transmission process of the WSN is presented and modeled using parallel-series configuration  of RBD\\
                        & \cite{bistouni2014analyzing} & The reliability of SEN is analyzed using RBD and the impact of switching stages on reliability is observed\\
                        & \cite{kanabar2009reliability} & The RBD-based reliability and availability of smart grid substation automation system is analyzed by considering various Ethernet switch architectures\\
\hline

Dependence & \cite{nojo1993incorporating}& The availability analysis of devices which are responsible for telecommunication, is performed using RBD technique In addition, different failure scenarios are also incorporated\\
                    & \cite{xin2013network} & RBD-based methodology is presented for the reliability analysis of general network systems\\
                    & \cite{wafler2013combined}& reliability analysis of complex communication network, that works inside the power grid system, is performed using RBD technique\\
\hline
Network Protocols  & \cite{zhou2007reecp} & A communication protocol is presented and modeled using series-parallel configuration of RBD\\
                    & \cite{damaso2014reliability} & An evaluation model is presented to incorporate power consumption\\
\hline
Network Topologies  & \cite{abd2014reliability} & The reliability of computer systems connected with each other through hypercube topology is analyzed using RBD technique\\
\hline
Dynamic Behavior & \cite{hai2012analysis}& A Bayesian network based RBD analysis of the WAP system, with dynamic behavior,  is presented for reliability and availability evaluation\\
                & \cite{torres1998bayesian}& A Bayesian network based RBD analysis of bridge network and power plant network, having dependant failure, is presented for reliability evaluation\\
                & \cite{shu2008wireless}& A fuzzy based RBD model of a sensor network is presented and its reliability is analyzed\\
\hline
Failure Prone Hardware/Software Components  & \cite{dharmaraja2012reliability} & The reliability analysis of VANETs system is performed by simulating its RBD model using \textit{SHARPE} software The RBD model presented in the paper also captures the characteristics of the network channel as well as the hardware failure scenarios\\
                                            & \cite{malek2008analytical}& The reliability and availability of an ICT-infrastructure system is analyzed by simulating its RBD model using \textit{Isographs} Reliability Workbench\\
                                            & \cite{palade2010reliability}& The reliability analysis of VoIP network is performed by simulating its RBD model using \textit{RAPTOR} tool\\
                                            & \cite{bobbio2010unavailability}& The availability analysis of SCADA communication links, which are integral part of a power grid station, is performed by simulating its RBD model using \textit{ITEM} software\\

\hline

Repairing Strategies& \cite{pokorni2011reliability}&  A methodology is presented for the reliability analysis of complex communication network using \textit{GPSSworld} simulation tool\\
                    & \cite{held2003consideration}&  A Monte Carlo simulation based reliability and availability analysis of long haul network is presented\\
\\ \bottomrule
%\end{tabular}
\end{longtable}

\begin{longtable}{p{0.25\linewidth}p{0.06\linewidth}p{0.6\linewidth}}
\caption{Representative summary of RBD-based reliability assessments using various \underline{\textit{Formal Methods}}}\label{RBD_formal}\\
%\begin{tabular}{p{3cm}p{1.5cm}p{12cm}}
\toprule
  \textbf{\emph{Network \newline Characteristics}} & \textbf{\emph{Ref.}}   &  \textbf{\emph{Brief summary}}\\
\midrule

 Fault Tolerance & \cite{guimaraes2011dependability} &  A stochastic Petri Nets based RBD analysis of enterprise network is performed for its reliability evaluation\\
                  & \cite{pereira2010dependability} & The reliability and availability analysis of enterprise network is performed by using RBD technique where the RBD models are solved through Petri Nets for analysis\\
\hline

Failure Prone Hardware/Software, Power Consumption, Safety & \cite{lira2013virtual}& The reliability and availability analysis of virtualized network resource allocation algorithm is performed through RBD modeling where the RBD model is then transformed into petri nets for analysis\\
  & \cite{fernandes2012dependability}& An RBD and SPN based methodology is presented for the reliability and availability analysis of virtualized network\\
  & \cite{wei2011dependability}& The reliability and availability analysis of virtual data center, which is an integral part of cloud computing, is performed using RBD technique and the RBD model is  converted into SPN for analysis\\
  & \cite{distefano2013evaluating}& The dynamic behavior of the sensors such as, sleep/wake policies are captured through DRBD and analyzed through Petri Nets for reliability evaluation of WSNs\\
  & \cite{omidi2012modeling}& The reliability and availability analysis of internet voting system is performed by modeling the system through RBD and the RBD model is then converted into SPN for analysis\\
  & \cite{lijie2012verification} & A safety communication protocol of EURORADIO train is modeled using RBD technique and analyzed using SPN\\
\\ \bottomrule
%\end{tabular}
\end{longtable}

\section{Reliability Analysis of Communication Networks with Fault Trees}
\label{sec:ReliabilityFaultTree}
Besides RBD, FT is an another widely utilized reliability analysis technique, which is mainly used to model the failure relationships among the communication system components and the effect of failure of components towards the overall communication system failure. In this section, we present a comprehensive survey of FT reliability analysis technique in the domain of communication networks. Similar to the RBD survey, in Section \ref{sec:AnalysisWithRBD}, this survey is further classified on the basis of analysis methods, i.e., analytical, simulation tools and formal methods.

\subsection{Analytical Methods with Fault Tree}

For a communication network to be functioning properly, it is essential that its \textit{critical components}, i.e., the components whose failure leads to the severe degradation in the performance of the whole communication network, are functioning properly. Fault Tree (FT) models allow us to observe the effects of these network component's failure on the overall reliability of the communication system. For example, the automated highways system (AHS), which is used to provide highest efficiency and safety to the railway transportation, consists of many critical components, such as data acquisition system, transmitter and communication system between successive vehicles. The impact of failure of these AHS components on the overall system failure has been modeled by using FT gates and analyzed analytically \cite{faghri1999application}. Similarly, the failure analysis of two schemes of transmission line protection system, which includes radial line protection scheme and Permissive Overreaching Transfer Trip (POTT) scheme, has been performed by modeling the critical components of the schemes using a FT \cite{schweitzer1997reliability}. Similarly, FT is also effectively used to observe the effects of a \textit{network topology} on the overall network reliability. For instance, the reliability of the Fiber Distributed-Data Interface (FDDI) token ring topology of a local area network (LAN) has been analyzed by using FT modeling \cite{bulka1992fault}.

Fuzzy logic and Bayesian networks based FTs have been commonly used to incorporate the effects of \textit{randomness, uncertainly and operational conditions} in the reliability analysis of communication systems. Some targeted areas of communication systems include communication control system \cite{peng2008approach}, Simple Networks Management Protocol (SNMP) \cite{prafiee}, Global Positioning System (GPS) \cite{song2009fuzzy}, intelligent substation's communication network \cite{wang2013analysis} and decentralized traction control system \cite{rong}. For instance, operational conditions that change over time and lead to random and uncertain faults have been captured by fuzzy logic based FTs \cite{peng2008approach} and thus the reliability of corresponding communication control system (CCS) is analyzed and optimized accordingly. Similarly, Bayesian networks have been used to simplify a large FT model of the intelligent substation for analysis purposes \cite{wang2013analysis}. Due to the inherent bidirectional reasoning technology, Bayesian networks have also been used to identify the weak links in the model with the aim to improve the reliability of the underlying communication network.

Beside communication networks, many interesting works have been found that utilized analytical approaches to analyze dynamic fault trees \cite{merle2011algebraic,cui2013minimal,merle2011dynamic,cheshmikhani2015probabilistic}. For instance, the authors present an algebraic framework to model dynamic gates and than can be used to determine the structure function of any Dynamic Fault Tree (DFT) \cite{merle2011algebraic}.

\subsection{Simulation Tools based Analysis with Fault Tree}

In many real-world communication networks, there are a large number \textit{critical communication components}, which makes the corresponding FT models quite complex and the corresponding paper and pencil based analysis become almost impossible. Therefore, computer simulation tools, such as  \textit{SHARPE} \cite{Sahner:1986:RMU:868535}, \textit{HARP} \cite{bechta1992dynamic} and \textit{OpenFTA} \cite{min2012fault}, have been widely adopted for the reliability analysis of communication networks. For instance, various critical components of a WSN, which includes base station, sensors, power supply, wireless channel, analog to digital converters (ADC), micro-controller etc., have been modeled using a FT and the failure analysis of a WSN is carried out using the SHARPE failure analysis tool \cite{kim2010hierarchical}. Moreover, \textit{SHARPE} has also been used to analyze the effects of the \textit{network topology} on the reliability of the communication networks \cite{silva2012reliability}. Similarly, Monte-Carlo simulation and the \textit{OpenFTA} tool have been used for the reliability analysis of Seamless Aeronautical Networking through the integration of Data links Radios and Antennas (SANDRA) demonstrator system and wide area protection system (WAP) \cite{ali2013sandra} \cite{dai2011reliability}. Moreover,  the HARP software has been frequently used for the FT based reliability analysis of \textit{fault-tolerant} mission avionic systems (MAS) and fault-tolerant hypercube systems \cite{bechta1992dynamic}.

Besides the above-mentioned fault tree simulation tools, there are many others that can provide many interesting  and powerfull features to analyze complex fault trees. Some of them include, \textit{MatCarloRE} \cite{manno2012matcarlore} that can solve hierarchical DFT, \textit{RAATSS} \cite{manno2012raatss} developed using Matlab toolbox for solving Repairable Dynamic Fault Tree, \textit{M{\"o}bius} \cite{daly2000mobius}, for general purposes performability evaluation based on Stochastic Activity Networks, \textit{OMNet++} \cite{varga2001omnet++}  provides a component architecture for models. Firstly, the components or modules are programmed in C++, then assembled into larger components  and \textit{Perfecto/UA} \cite{cavalieri2010performance} gives a library for the performance evaluation in OPC UA Communication based on OMNet++.

Many approaches have been proposed in order to reduce the computational efforts of simulation approaches \cite{chiacchio2013weibull,chiacchio2011open,manno2014conception,yevkin2015efficient}. For instance, a composition algorithm is used to generalize the traditional hierarchical technique that is able to reduce the computational effort of a large DFT \cite{chiacchio2013weibull}. Similarly, DFT reliability solvers are developed, based on the Monte Carlo simulative approach, and are written using Matlab® code \cite{chiacchio2011open,manno2014conception}. Approximate Markov chain method for dynamic fault tree analysis is proposed for both reparable and non-reparable systems that readily improve the analysis capabilities \cite{yevkin2015efficient}.

\subsection{Formal Models with Fault Tree}

The \textit{COMPASS} tool \cite{bozzano2009compass}, developed at the RWTH Aachen University in collaboration with the European Space Agency (ESA), supports the formal FT analysis, specifically for aerospace systems. For verification purposes, \textit{COMPASS} provides support of several model checking tools, like the \textit{NuSMV} \cite{cimatti2002nusmv} and \textit{MRMC} \cite{katoen2005Markov} model checkers. This tool provide various templates containing placeholders that have to be filled in by the user. These templates are primarily composed of the most frequently used patterns that allow easy specifications of property by non-experts by hiding the details of the underlying temporal logic. The tool generates several outputs, such as traces, FTs and FMEA tables, along with diagnostic and performance measures. However, to the best of our knowledge, the tool has not been used in the context of verifying communication networks

Some efforts have been made to analyze the FT formally using theorem proving techniques. For example, the Interval Temporal Logic (ITS), i.e., a temporal logic which is a combination of first order and propositional logic, based FT tree analysis is presented in \cite{hansen1998safety,ortmeier2007formal}. The work in \cite{hansen1998safety} provides the formalization for a rail-road crossing case study and the logic gates of FT are modeled using propositional and temporal logic and the verification of the overall FT is carried out by using the Karlsruhe Interactive Verifier (KIV) \cite{reif1992kiv} system. Although, the analysis is formal but the fault tree construction is intuitive. Similarly, the work presented in \cite{xiang2004fault}, proposes to first construct the FT formally by using the Observational Transition Systems (OTS) \cite{huisman2006temporal} and then conduct the formal analysis by using CafeOBJ \cite{azvan1998cafeobj,azvan1998cafeobj}, which is a formal specification language that verifies software and hardware systems interactively. Moreover, a higher-order-logic formalization of FT gates has been recently proposed in \cite{WAhmad_CICM15,ahmed2016FT,ahmed2016formal}, which allows conducting the FT analysis within the sound core of a higher-order-logic theorem prover. However, none of the above-mentioned works have been used to conduct the formal dependability analysis in the domain of communication networks.

The \textit{dynamic behavior} of networks components, such as timed behavioral nature, cannot be captured by simple FT models but Petri Nets provide a very feasible alternative for this purpose. The communication network under consideration is modeled with a FT, which is then transformed into its corresponding Petri Nets based model for analysis. For example, the reliability of the broadband integrated service network (B-ISDN) has been assessed by modeling the dynamic re-routing mechanism of the traffic by using the FT-based Petri Net approach \cite{balakrishnan1996stochastic}.

A summary of reliability works using FT-based modeling is presented in Table \ref{FT}.

\begin{longtable}{p{0.25\linewidth}p{0.06\linewidth}p{0.6\linewidth}}
\caption{Representative summary of Fault Tree reliability assessments using various \underline{\textit{Analytical/Simulation/Formal Methods}} analysis techniques}\label{FT}\\
%\begin{tabular}{p{3cm}p{1.2cm}p{12cm}}
\toprule
 \textbf{\emph{Network \newline Characteristic}} & \textbf{\emph{Ref.}}   &  \textbf{\emph{Brief summary}}\\
\midrule
Failure Prone Hardware/ \newline Software Components & \cite{faghri1999application} & A longitudinal control system of the automated highway system is modeled and thus its failure probability is calculated\\
                                           & \cite{schweitzer1997reliability}& A transmission line protection system is analyzed in this paper\\
                                           & \cite{bulka1992fault} & A communication topology of LAN, named FDDI token ring, is modeled and analyzed in this paper\\
\hline
Dynamic Behavior & \cite{peng2008approach}& Due to the randomness and fuzziness of CCS system, its behavior is captured by fuzzy fault tree for reliability analysis\\
               & \cite{prafiee}& An SNMP protocol based DCN system, possessing fuzziness is modeled and analyzed in this paper\\
               & \cite{song2009fuzzy}&  A system is first modeled through FT and then FT model is transformed into fuzzy fault tree to capture fuzziness of the system\\
               & \cite{wang2013analysis}& A star topology based DCN of intelligent substation is modeled and analyzed\\
               & \cite{rong}& A decentralized traction control system of the communication network is modeled\\
             &  \cite{balakrishnan1996stochastic} & A B-ISDN network deployed in a power plant, is modeled and analyzed using Petri Net approach\\

\hline
Dependent Failures/ \newline Topologies/ \newline Components Effects&  \cite{kim2010hierarchical,silva2012reliability}& The wireless sensors connected with each others through different communication topologies, and a cluster based WSN are modeled and analyzed using \textit{SHARPE} simulation tool\\
                                                & \cite{dai2011reliability}& The communication layers such as, regional communication, substation communication and wide area communication of the communication system of WAP are modeled and analyzed using \textit{OpenFTA}\\
                                                & \cite{ali2013sandra}& A SANDRA demonstrator system of the avionics system is analyzed using \textit{OpenFTA}\\
\hline
Fault Tolerance&  \cite{bechta1992dynamic}& A fault tolerant MAS system, with redundancy approach and recovery approach is modeled and analyzed using \textit{HARP}\\
\bottomrule
%\end{tabular}
\end{longtable}

\section{Reliability Analysis of Communication Networks with Markov Chains}
\label{sec:ReliabilityMC}
Markov chains (MCs) offer a state-based mathematical modeling technique, which has been extensively utilized to capture the dynamic and probabilistic behavior of communication networks. The analysis start by constructing a state based model of a given network system, known as a Markovian model, which is then analyzed either analytically, or using simulation or formal methods based techniques. Reliability analysis using MC is amongst the most commonly used approaches and have been used in various communication domains, including client server networks, ad-hoc networks, WSNs, high-speed train networks and network storage systems. In this section, we present a comprehensive survey of these contributions, while classifying them according to the underlying analysis techniques like in the previous two sections.

\subsection{Analytical Methods with Markov Chain Based Analysis}

Due to rapidly increasing usage of communication services in our daily life, the problems related to \textit{data communication} is also becoming dynamic and adverse, which consequently affects the wireless channel unavailability, transmission line failure and latency delays in data transmission. To design a network that can mitigate these affects, the reliability analysis of the network is indispensable. It has been found that MC is the most widely used reliability modeling technique to incorporate the dynamic behavior of the data transmission system in a communication  network for reliability analysis. Some of the major domains of communication networks that utilize MC for reliability analysis include wireless ad-hoc networks \cite{khabazian2013performance} \cite{tan2011analytical} \cite{chen2002network}, communication machine 2 (CM-2) \cite{ahluwalia1992performance},  multi-hop linear networks \cite{hassan2011quasi}, high speed trains \cite{junfeng2001analysis} \cite{miao2013performance}, multi-server client server systems \cite{lin2011markov}, mobile cloud computing systems \cite{park2011markov} \cite{shi2013framework} \cite{routaib2014modeling}, cloud storage systems \cite{zhang2013modeling}, cellular networks \cite{jindal2011markov} \cite{venmani20123ris}, wide area measurement (WAM) systems  \cite{wen2009reliability}, cognitive radios networks \cite{balapuwaduge2014system} and  emergency communication systems \cite{wolff2011performance}. Considering a vehicular ad-hoc network \cite{khabazian2013performance}, which is considered to be unreliable due to lack of acknowledgment of the broadcasted messages by the receiving vehicles, the varying number of broadcasted messages are modeled with a 2D MC to present the dynamic message transmission process of the VANETs. The MC is then analyzed to evaluate the reliability of the communication network in terms of messages transmission delays.

\begin{longtable}{p{0.25\linewidth}p{0.06\linewidth}p{0.6\linewidth}}
\caption{Representative summary of Markov-based reliability assessments using various \underline{\textit{Analytical/Simulation}} analysis techniques}\label{Mar_Ana}\\
%\begin{tabular}{p{3cm}p{1.5cm}p{12cm}}
\toprule
 \textbf{\emph{Network \newline Characteristic}} & \textbf{\emph{Ref.}}   &  \textbf{\emph{Brief summary}}\\
\midrule
Data Communication System  & \cite{khabazian2013performance}& The performance in terms of transmission delay is evaluated for safety message broadcast system through Markov Chain\\
                           &\cite{tan2011analytical}& The performance in terms of wireless communication is evaluated for drive through internet system through Markov Chain\\
                           & \cite{chen2002network}& A Markov chain based quantitative approach has been proposed for the reliability analysis of WSNs\\
                           & \cite{ahluwalia1992performance}& The performance in terms of message delay is evaluated for the communication machine 2 (CM-2) through DTMC modeling\\
                           & \cite{hassan2011quasi}& The reliability of multi hop linear network is analyzed using quasi stationary Markov Chain\\
                           & \cite{junfeng2001analysis}& The reliability of wireless data transmission system of high speed rail system is analyzed using Markov Chain\\
                           &\cite{miao2013performance}& The reliability of high speed train is analyzed using Markov Chain\\
                           & \cite{lin2011markov}& The reliability of multi server client system is analyzed using Markov Chain\\
                           & \cite{park2011markov}& A strategy is proposed of improving the reliability of mobile computing system using Markov Chain. the proposed strategy encounters mobiles volatility\\
                           & \cite{shi2013framework}& A framework is proposed to improve the service selection system of the cloud computing system and its performance is analyzed using Markov Chain\\
                           & \cite{routaib2014modeling}& The performance in terms of latency delay is analyzed of the cloudlet based centralized architecture for mobile cloud computing system using CTMC\\
                           & \cite{zhang2013modeling}& A Markov Chain based reliability analysis framework is presented to improve the replication techniques in cloud storage system\\
                           & \cite{jindal2011markov}& The survivability in terms of call blocking and excess delay due to failure is analyzed for cellular networks using Markov Chain\\
                           & \cite{venmani20123ris}& A novel approach has been proposed for reducing cost and improving availability of 4G-LTE mobile networks using Markov Chain\\
                           & \cite{wen2009reliability}& A Markov Chain based reliability analysis of wide area measurement (WAM) system is presented\\
                           & \cite{wolff2011performance}& The reliability analysis of emergency communication system is performed using Markov Chain\\
                           & \cite{balapuwaduge2014system}& Availability analysis of the wireless channel of cognitive network radios (CNR) is performed using CTMC\\
\hline
 Fault Tolerance & \cite{taqieddin2011survivability}& A Markov Chain based reliability and availability of the trusted two link protocols of the WSNs is performed\\
                & \cite{kumar2011reliability} & The reliability of fault tolerant WSN is analyzed using Markov Chain\\
                & \cite{rao2011reliability}& A fault tolerant network storage system is presented and some redundancy mechanisms are proposed. Each redundancy mechanism of the system is analyzed and compared in terms of reliability\\
                & \cite{daoudperformance}& The reliability and availability analysis of gigabit Ethernet network control system is performed using CTMC\\
                & \cite{zimmerling2013modeling}& The reliability in  terms of synchronous transmission is analyzed for multi-hop system using Markov Chain\\
                & \cite{xiao2011performance}& A Markov Chain based reliability analysis of a power communication system such as spider web, is presented\\
\hline
 Channel Characteristics & \cite{5601392}& The reliability in terms of wireless channel communication and link quality is analyzed for WSN using Markov Chain\\
                         & \cite{fort2013availability}& The reliability of global system for mobile communication rail (GSM-R) is analyzed using Markov Chain\\
                         & \cite{zhu2012service}& The reliability of copper and fiber based architecture of railway communication system is analyzed using Markov Chain\\
                         & \cite{ciufudean2007reliability}& The reliability analysis  of randomized pulse modulation scheme of a security communication system has been presented\\
\hline
 Dependant Failures & \cite{xuejie2013reliability}& The reliability and availability in terms of service faults and physical breakdown is analyzed for cloud computer service using a Markov Chain and as well as MTTR/MTTF formulas\\
                    & \cite{dantas}& The availability analysis of Eucalyptus cloud computing is presented\\
                    & \cite{chen2005reliability}& A wireless network system such as, wireless CORBA system with imperfect network components, is analyzed to calculate end-to-end instantaneous reliability\\
                    & \cite{cui2013methods}& A Markov Chain computationally inexpensive approach has been presented for the reliability analysis of cloud services with dependent failures\\
\bottomrule
%\end{tabular}
\end{longtable}

\begin{longtable}{p{0.25\linewidth}p{0.06\linewidth}p{0.6\linewidth}}
\caption{Representative summary of Markov-based reliability assessments using various \underline{\textit{Formal Methods}}}\label{Mar_Ana_Sim_Formal}\\
%\begin{tabular}{p{3cm}p{1.5cm}p{12cm}}
\toprule
 \textbf{\emph{Network \newline Characteristic}} & \textbf{\emph{Ref.}}   &  \textbf{\emph{Brief summary}}\\
\midrule
Behavioral Networks/
Uncertainties &  \cite{blake1989reliability}& A Markov Chain based reliability analysis approach is presented for MINs. The approach first tends to evaluate MINs at component level and finally at system level\\
              & \cite{kamyod2012resilience} \cite{kamyod2013end}& The reliability and availability of IP multi media subsystem is analyzed using Markov Chain\\
              & \cite{kamyod2012resilience} \cite{kamyod2013end}& The reliability and availability of IP multi media subsystem is analyzed using Markov Chain\\
              & \cite{Matos2014}& The availability of mobile cloud computing system is analyzed by modeling the physical components using CTMC\\
              & \cite{wafler2013combined}& A Markov Chain based approach is presented to analyze the reliability of smart grid station having dependent failures\\
              & \cite{navarro2007predictive}& An MS-exchange email server having uncertain failure rates and repair rates is modeled and analyzed\\
              & \cite{dantas}& The availability analysis of Eucalyptus cloud computing is presented\\
              & \cite{Matos2014}& The availability of mobile cloud computing system is analyzed by modeling the physical components using CTMC\\
\hline
Data Communication Networks/ \newline
Concurrent Failutes & \cite{dharmaraja2012reliability} & The survivability analysis of VADHOC network, in terms of message lost, is modeled using Markov reward modeling and analysis is done using \textit{SHARPE} software\\
                     &\cite{munir2011markov}& A reliable WSN system is presented for safety applications and its reliability analysis is done using Markov Chain and analysis is done in \textit{SHARPE } software\\
                     & \cite{palade2010reliability}& The reliability analysis of VoIP network is performed by simulating its RBD model using \textit{RAPTOR} tool\\
                     & \cite{engelen2014reliability}& The reliability and availability analysis of nanosatellites based spacecrafts-swarm is performed using Markov Chain and analysis is done through Monte Carlo simulation\\
\hline
 Data Communication Network/ \newline
Network Protocols & \cite{conghua2013analysis}& The reliability and rapidness analysis of fast and secure protocol, in term of successful delivery and time elapsed, is performed using Markov Chain and analysis is done in \textit{PRISM} model checker\\
                  & \cite{petridou2013survivability}& The survivability analysis of WSN, in terms of  data loss and frequency of failures, is performed using PRISM model checker\\
                  & \cite{massink2004model}& The dependability properties of wireless group communication are verifies using \textit{ETMCCC} model checker\\
\hline
Data Communication System/ \newline
Unreliable Components& \cite{gharbi2011algorithmic}& A finite retrial system with unreliable and multi servers, is analyzed through Petri Nets\\
                     & \cite{zhu2012service}& WLAN based data communication systems are presented for CTBC and are analyzed through Petri Nets\\
                     & \cite{jindal2011markov}& The survivability analysis of cellular network, in terms of excess delay due to failure and call blocking, is performed using Petri Nets\\
                     & \cite{schoenen2013erlang} \cite{zhang2013performance}& Erlang-B and Erlang-C traffic scenarios are created and thus reliability is evaluated\\
                     & \cite{zeng2009spn}& WSNs are modeled and analyzed through SPN to observe the effects of data packet number, throughput and arrival rate\\
                     &\cite{christodoulou1994petri}& A timed rotation token protocol is analyzed for the reliability analysis of FDDI using Petri Nets\\
                     & \cite{ibe1993performance}& Two types of networks architectures are modeled for a file server system and are analyzed using Petri Nets\\
                     & \cite{279721}& The reliability analysis of a finance service network is performed using Petri Nets\\
                     & \cite{sun2009survivability}& The reliability analysis of a system with three computers and distributed memory is performed using Petri Nets\\
                     & \cite{zeng2011spn}& A system with one user and two users is analyzed using Petri Nets\\
\\ \bottomrule
%\end{tabular}
\end{longtable}

Due to the ability of a MC to incorporate \textit{components redundancy}, it has been used to evaluate the reliability of the networks that are comprised of redundant components, like wireless ad-hoc networks \cite{taqieddin2011survivability}, WSNs \cite{kumar2011reliability}, network storage systems \cite{rao2011reliability} and gigabyte Ethernet systems \cite{daoudperformance}. For instance, in a wireless ad-hoc network whose functionality is defined by the Trust Levels Routing Protocol (TLR), the wireless transmission nodes of the network are redundant \cite{taqieddin2011survivability}. The behavior of these redundant nodes is modeled and analyzed with a finite state CTMC for the reliability analysis of the ad-hoc network. Moreover, MCs have also been used to analyze the reliability of \textit{energy efficient communications networks}, which include low power wireless multi-hop networks \cite{zimmerling2013modeling} and power line communication networks \cite{xiao2011performance}. For instance in a low-power multi-hop network, the functionality of the Synchronous Protocol (ST), which greatly reduces the overall network energy consumption and simplifies the mathematical modeling of the network, has been modeled and analyzed by MCs \cite{zimmerling2013modeling}.

\textit{Mobility} of the network components considerably affects the reliability and availability of any communication network as the behavior of the network becomes dynamic due to the changing position of the network components. Due to the dynamic characteristic of MCs, they have been widely used to capture the effect of component's mobility on the overall reliability and availability of the communication networks. Some of the reliability analysis work in this context includes industrial WSNs \cite{5601392}, high speed train communication networks \cite{fort2013availability} \cite{zhu2012service}, and  randomized pulse modulation (RPM) \cite{ciufudean2007reliability}. Consider the example of industrial WSNs, where the dynamic behavior of the sensors, which is mainly due to their mobility and harsh environment in an industrial area, has been modeled with a finite-state MC. This MC is then solved to evaluate the end-to-end communication reliability of the WSN \cite{5601392}. Besides mobility, the effect of network \textit{components failures} on the overall network's reliability and availability have also been widely observed by modeling the communication network with MCs, where the components failures are defined by some unique Markov states. For instance,  the effect of service failures and resource breakdowns on the reliability of cloud computing network has been observed using MCs in \cite{xuejie2013reliability} \cite{dantas}. Some other relevant works in this regard include the reliability analysis of cloud services \cite{cui2013methods} and wireless network CORBA \cite{chen2005reliability}.

The size of the MC significantly increases with the increase in the size of the communication network. For large communication networks, MC-based reliability analysis become challenging due to the exponential increase in the number of states. A hybrid approach, i.e., a combination of RBD and MC, has been used to cater this issue and thus conduct the reliability analysis of large communication networks,  such as mobile cloud computing system \cite{Matos2014}, cloud computing system \cite{dantas}, MINs \cite{blake1989reliability}, IP multimedia subsystems \cite{kamyod2012resilience} \cite{kamyod2013end} and power grid communication systems \cite{wafler2013combined}. For a particular communication network, the components that possess timed behavioral nature are modeled with MCs whereas the other components are modeled with RBDs. For example,  the components of the cloud computing, including Wi-Fi, battery and resource mobility, have been modeled with RBD whereas the MC has been developed to capture the timed behavioral nature of the network in \cite{Matos2014}. Moreover, communication networks with uncertain workload parameters, imprecision, and imperfect coverage have been analyzed with fuzzy logic based MCs \cite{navarro2007predictive}. The network is modeled with MC to present the probabilistic behavior and the model is enhanced with fuzzy logic to deal with the uncertainties. For instance, both MC and fuzzy logic have been utilized to conduct the reliability analysis of disk arrays, which are used to store the data of MSExchange like email servers in \cite{navarro2007predictive}.

\subsection{Simulation Tools based Analysis with Markov Chain}

As described above, the MC becomes quite large for communication networks with complex behaviors and too many components. Considering the example of fault tolerant WSNs, the total number of sensor nodes in a wireless cluster doubles when one extra spare sensor node is added with each sensor node \cite{munir2011markov}. Due to the increase in the total number of sensor nodes, the complexity of the network significantly increases and this yields to a significantly larger MC.  Specialized simulation tools for MC based reliability analysis tackle these problems quite well. The \textit{SHARPE} tool has been used for the reliability analysis of VANETs \cite{dharmaraja2012reliability} and fault tolerant WSNs \cite{munir2011markov}. Similarly, the \textit{RAPTOR} tool has been used for the analysis of VoIP communication system \cite{palade2010reliability} and Monte Carlo simulation based reliability analysis of satellite swarms \cite{engelen2014reliability}.

\subsection{Formal Methods with Markov Chain}

Model Checking \cite{alur1990model} is one of the powerful formal methods that facilitates the user to describe the behavior of a given system in the form of a state machine and verify the user-defined properties against it. It verifies the properties by checking each state rigorously and provide accurate results. Probabilistic Model Checking technique extends the traditional model checking principles for the analysis of MCs and allows the verification of probabilistic properties. Some notable probabilistic model checking tools include \textit{PRISM} \cite{lin2010modeling} and \textit{ETMCC} \cite{hermanns2003etmcc}. Probabilistic Model Checking technique has been considerably adopted to verify the reliability and availability properties of many communication networks, such as Fast And Secure Protocol (FASP) \cite{conghua2013analysis}, WSNs \cite{petridou2013survivability} and wireless group communication \cite{massink2004model}. Considering the reliability analysis of FASP, the reliability of successful data transmission is defined in Stochastic Temporal Logic (STL) whereas the communication network is modeled in the form of a sender, receiver and a communication channel module in \textit{PRISM}. The reliability property is then verified against the communication network using the \textit{PRISM} model checker.

Recently, a higher-order-logic formalization of time-homogeneous DTMC with finite state space has been presented using the HOL theorem prover \cite{liu2013formal}. This foundational formalization has been used to verify some simple case studies, including a simple binary communication channel.

A lot of work has been done on analyzing reliability of communication networks using Petri Nets with Markov chains.  The main idea is to first model the system using MC and then transform it into its corresponding Petri Net model, which is finally analyzed. Using this approach, the reliability analysis of finite source retrial system that possesses unreliable hardware and multi servers is presented in \cite{gharbi2011algorithmic}. Some other prominent works in this direction include the reliability analysis of the data communication systems of the WLAN based train control system \cite{zhu2012service}, cellular networks \cite{jindal2011markov} and WSNs \cite{schoenen2013erlang}  \cite{zhang2013performance} \cite{zeng2009spn}. Moreover, some network protocols, like  courier and FDDI token ring protocol, have also been analyzed using the Petri Net approach \cite{youness2006robust} \cite{christodoulou1994petri}. The effects of the network architecture on the reliability of the whole system is presented for the file server system \cite{ibe1993performance} , finance service networks \cite{279721}, distributed memory network    \cite{sun2009survivability} and Low Earth Orbit (LEO) satellite network \cite{zeng2011spn}.

A summary of reliability works using MC modeling with analysis performed through analytical/simulation and formal methods are presented in Tables \ref{Mar_Ana} and \ref{Mar_Ana_Sim_Formal}, respectively.

\subsection{Network Metrics}

RBD and FT have not been directly utilized for network metrics assessment due to the fact that they are especially developed for reliability analysis.
%They have been used indirectly to improve the network metrics we do not consider them to fall into the present scope of the paper.
On the other hand, MC is a flexible technique and has been used for analyzing a variety of network metrics for instance, downtime, throughput and resilience. Some key works in this direction are listed in Table \ref{table:MC_net_metr}.
%The main focus of this paper is to gather and then to give a comprehensive comparison among the works that have been done related to reliability analysis in the area of communication networks by using RBD, FT and MC. However, we have presented a summary of works table that that utilizes MC for network metrics assessment in Table \ref{table:MC_net_metr}  as follows:

\begin{longtable}{p{0.25\linewidth}|p{0.08\linewidth}|p{0.6\linewidth}}
\caption{Representative summary of Communication Networks Metrics Assessment Using Markov Chains}\label{table:MC_net_metr}\\
%\begin{tabular}{|p{0.25\linewidth}|p{0.08\linewidth}|p{0.6\linewidth}|}
  \hline
  Network Metrics & Ref. & Description \\
  \hline
  \hline
  Throughput &  \cite{tahir2008markov}  &  Markov chains have been used to analyze the carrier sense multiple access (CSMA) type MAC
protocol for its delay and throughput characteristics with and without transmitter power control\\
                &  \cite{javid2016analysis}     &  Analyzed the achievable region of the throughput in a slotted ALOHA-based two-way relay network by using Markov chains\\
                & \cite{chang2008markov}      & Analyzed the number of polls (NPS) in Wimax networks based on
the Markov chain analysis for minimizing the average polling delay
while increasing network throughput\\
\hline
  Latency & \cite{vellambi2011throughput}  & Using Markov chains, the authors investigate the effect of finite buffer sizes on the throughput capacity and latency in terms of packet delay of line network with packet erasure links that have perfect feedback  \\
    & \cite{torabkhani2012throughput} & A Markov chain method has been proposed in order to estimate the throughput
and average latency in wireless erasure networks with nodes having finite buffers\\
\hline
  Resilience & \cite{cholda2006reliability} & Reliability Assessment of resilient packet ring (RPR) is carried out using Markov chains   \\
             &  \cite{martinez2008behavior}  & Error resilience parameters are proposed for Mobile AdHoc Network (MANET) network scenarios as an
appropriate error resilience configuration, which are then analyzed using Markov chains\\
            &\cite{wang2003markov} & Using Markov chains, the authors systematically analyze the features of resilience to failures and attacks of the current structured P2P systems in terms of average path length and hit ratio and understand the causes, which lead to better resilience features. \\
            & \cite{rusekeffective} & The author discusses business impact analysis in the context of resilient
communication networks. It is based on the total (aggregated) penalty that may be
paid by an operator when the services (identified with transport demands) provided
are interrupted due to network failures. First a CTMC model is constructed and then means and variances of compensation policy-related penalty
values for all the services are found using Markov chains. \\
 \hline
  Producibility & \cite{bruneo2010dependability} & A new dependability
parameter is defined, referred to as producibility, which is able to capture the
capability of a sensor to accomplish its mission and is estimated by using Markov chains.   \\
  \bottomrule
\end{longtable}

\section{Insights and Common Pitfalls}
\label{sec:InsightsPitfalls}

In this section, we endeavor to clarify the difference among the reliability modeling techniques of Section \ref{sec:depend_model} by comparing them and highlighting their advantages and limitations. The comparison among these modeling techniques is shown in Table \ref{RFM_comp}.

The criteria for the selection of these modeling techniques, for a certain system, mainly depends upon the type of system and problem domain. For instance, RBD is primarily used if we are interested in the successful working of the system while the FT models the failure relationship due to the failure of individual component of the system. Although, both of these techniques work well for combinatorial type of problems but cannot handle the non-combinatorial problems. On the other hand, Markov chain provides analysis for a wide variety of problems but fails to cater for large and complex systems due to exponential growth in the number of states.

\begin{figure}[!ht]
  \centering
  \includegraphics[width=\linewidth,trim = {1.2cm 0cm 0cm 0cm}, clip]{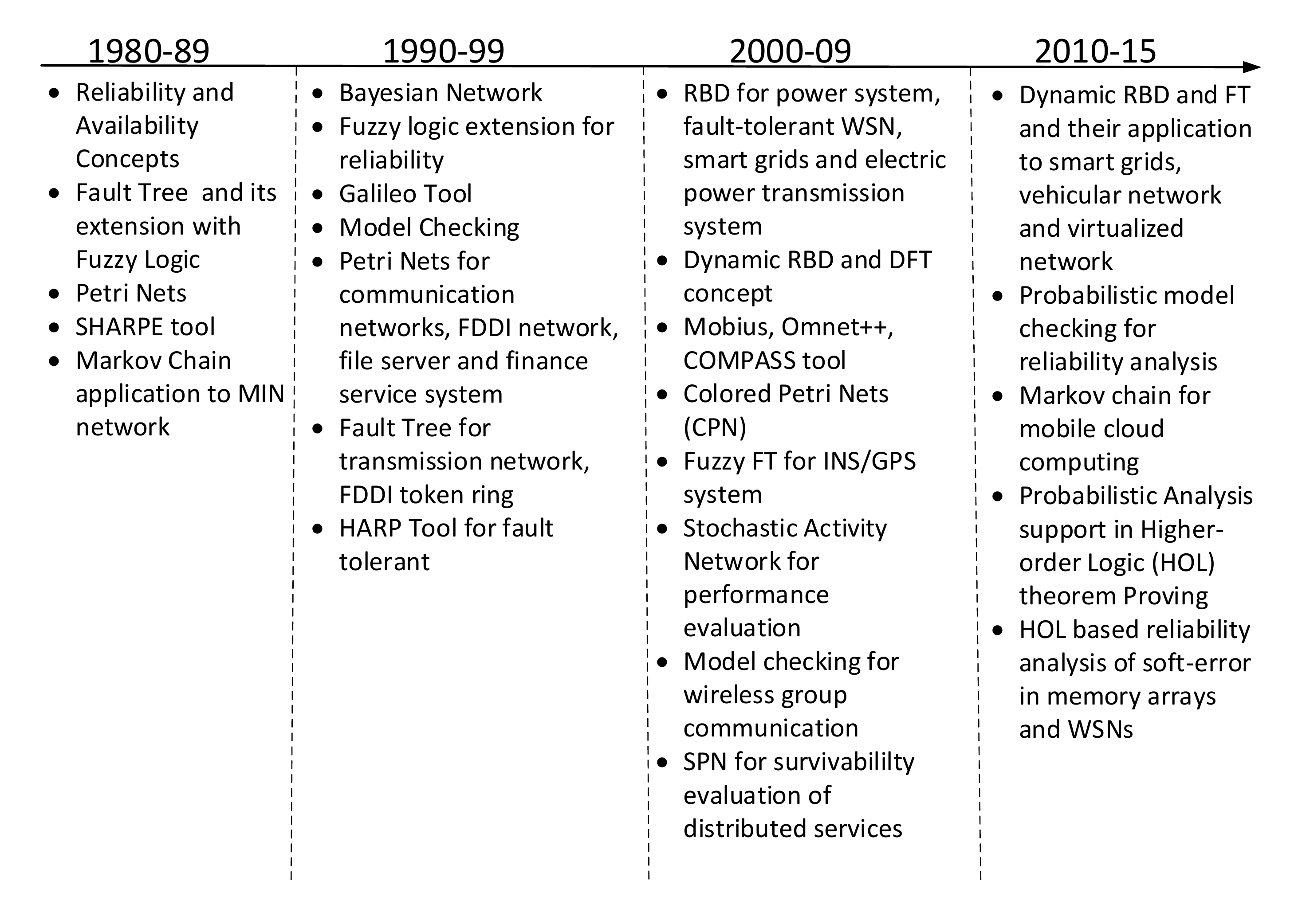}
  \caption{Timeline of the Surveyed Techniques}\label{fig:timeline_survey}
\end{figure}

An abstract overview of the trends of surveyed works in this paper is presented, as shown in Figure \ref{fig:timeline_survey}. It has been observed that the major focus in  1980s and 90s was on the development of primitive reliability modeling techniques. Some of these modeling techniques had been utilized for the communication network problems. However, in 2000-09, the major focus moved to the amalgamation of different modeling techniques, development of new simulation tools for reliability assessment and utilization of formal methods, such as Petri Nets and model checking, in the domain of reliability analysis. It has been noticed from the last five years that one of the active research areas is the incorporation of capturing dynamic behaviors in RBDs and FTs and their application to communication network problems. A significant development has been also done in developing the foundational framework for reliability assessment using HOL theorem proving \cite{WAhmad_CICM14,WAhmad_CICM15,WAhmed_IWIL15} and it has been successfully utilized for the reliability analysis of WSN data transport protocols \cite{WAhmed_Wimob15}.

\begin{figure}[!ht]
  \centering
  \includegraphics[width=\linewidth,trim = {0cm 1cm 0cm 0cm}, clip]{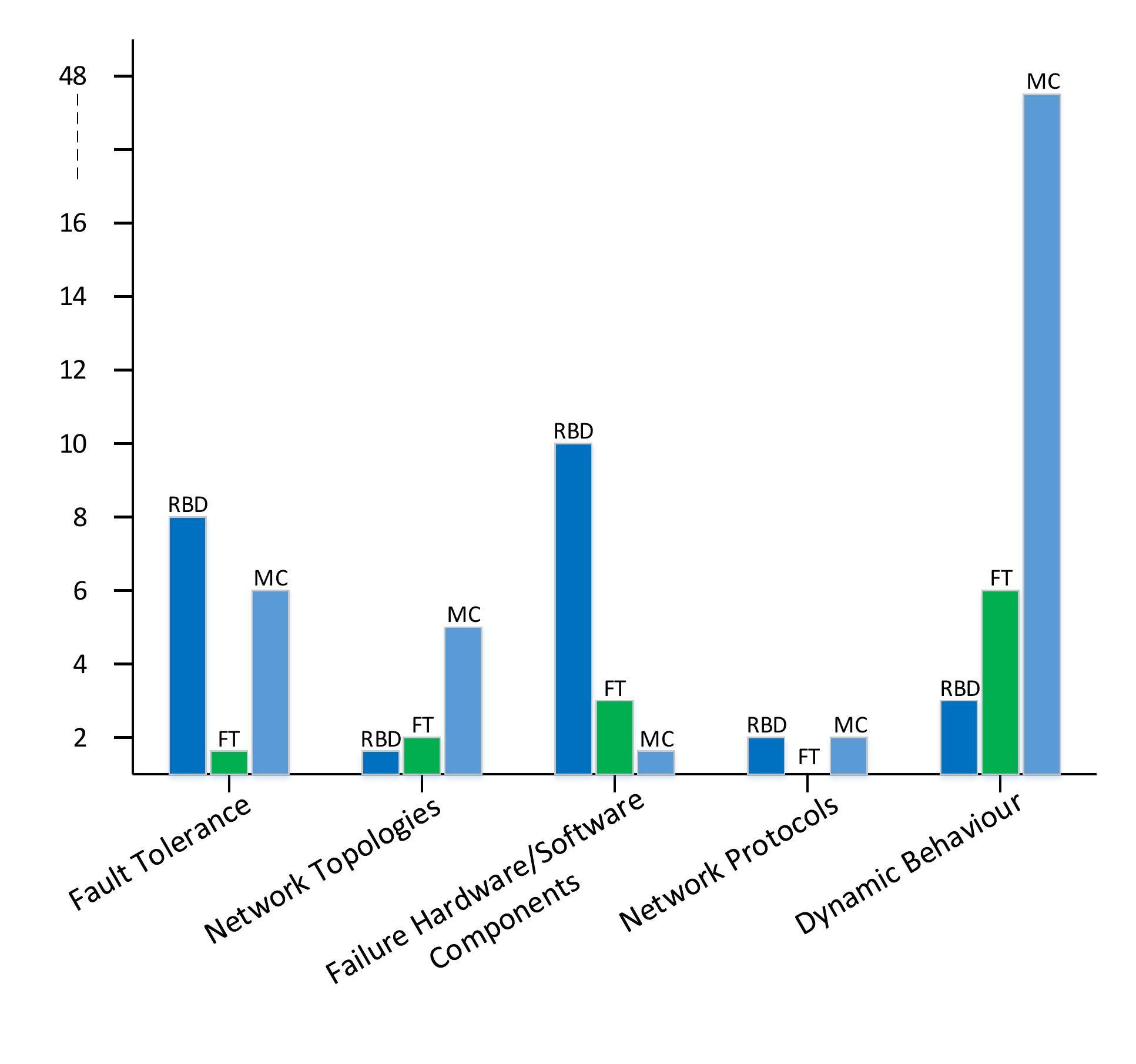}
  \caption{Histogram Showing Number of Papers for each Technique }\label{fig:histo_survey}
\end{figure}

An histogram overview of network characteristics versus number of surveyed paper with respect to reliability modeling techniques is shown in Figure \ref{fig:histo_survey}. The number of papers for MC is much more for analyzing dynamic behaviours compared to RBD and FT.

Although, MC is the most utilized reliability modeling technique for network problems but we can notice from Table \ref{network_charac_rel_be} that few researchers utilized MC in the area of \textit{network topologies} and \textit{network protocols}. On the other hand, FTs and RBDs have been extensively utilized for \textit{fault tolerance} and \textit{software/hardware failures} related network reliability problems. But, not much work has been done using these techniques for modeling and analyzing \textit{message delay} and \textit{channel characteristics} network problems. Similarly, simulation methods have been used for almost every network characteristic and among formal methods, Petri Nets have been the most utilized method, especially for \textit{software/hardware} related reliability analysis network problems.

\begin{center}
\begin{longtable}{p{0.25\linewidth}| p{0.08\linewidth} |p{0.08\linewidth} |p{0.08\linewidth} |p{0.2\linewidth} |p{0.2\linewidth}}
\caption{Representative summary of Reliability Modeling and Analysis Techniques used for Communication Network Characteristics}\label{network_charac_rel_be}\\
%\begin{table}
%  \begin{flushleft}
%\begin{tabular}{|m{0.25\linewidth}|m{0.08\linewidth}|m{0.08\linewidth}|m{0.08\linewidth}|m{0.2\linewidth}|m{0.1\linewidth}|}
%\begin{tabularx}{1\linewidth}{XXXXXXXX}
\hline
  \textbf{\emph Network Behaviour} & \textbf{\emph FT} & \textbf{\emph RBD}  & \textbf{\emph MC} & \textbf{\emph Simulation} & \textbf{\emph Formal Methods} \\
  \hline
  Fault Tolerance & \cite{bechta1992dynamic} &  \cite{bein2005reliability,ghaffari2008fault,bistouni2014pars,jiajia,yang2009impact,chen2009design}&    \cite{taqieddin2011survivability,kumar2011reliability,rao2011reliability,daoudperformance,zimmerling2013modeling,xiao2011performance} & \cite{taqieddin2011survivability,kumar2011reliability,rao2011reliability,daoudperformance,zimmerling2013modeling,xiao2011performance} & \cite{guimaraes2011dependability,pereira2010dependability}  \\
  \hline
  Network Protocols & \cite{prafiee}  & \cite{zhou2007reecp,damaso2014reliability}   & \cite{conghua2013analysis} & \cite{prafiee} & \cite{WAhmed_Wimob15} \cite{conghua2013analysis}  \\
  \hline
 Failure Prone Hardware/Software Components & \cite{faghri1999application,schweitzer1997reliability,bulka1992fault} & \cite{dharmaraja2012reliability,malek2008analytical,palade2010reliability,bobbio2010unavailability}   & \cite{xuejie2013reliability} & \cite{faghri1999application,schweitzer1997reliability,bulka1992fault} \cite{xuejie2013reliability} &  \cite{lira2013virtual,fernandes2012dependability,wei2011dependability,distefano2013evaluating,omidi2012modeling,lijie2012verification} \\
 \hline
  Message Delay &  & &   \cite{jindal2011markov,khabazian2013performance,routaib2014modeling} &  \cite{jindal2011markov,khabazian2013performance,routaib2014modeling} &   \\
  \hline
  Channel Characteristics &  & \cite{dharmaraja2012reliability}    & \cite{5601392,fort2013availability,zhu2012service,ciufudean2007reliability} & \cite{5601392,fort2013availability,zhu2012service,ciufudean2007reliability} \cite{dharmaraja2012reliability}  &   \\
  \hline
  Network Topologies & \cite{wang2013analysis,bulka1992fault,kim2010hierarchical,silva2012reliability} & \cite{abd2014reliability}   &  & \cite{wang2013analysis,bulka1992fault,kim2010hierarchical,silva2012reliability} & \cite{WAhmed_Wimob15}  \\
  \bottomrule
\end{longtable}
\end{center}

\subsection{Comparison of Reliability Modeling Techniques}

In this section, we will compare and contrast the various reliability modeling techniques presented in this paper. In particular, we will provide a discussion on the advantages and limitations of reliability modeling using the techniques of RBDs, FTs, MC, and BNs, respectively. A tabulated summary of this comparison is presented in Table \ref{RFM_comp}.

\subsubsection{Advantages and Limitations of Reliability Block Diagrams}

\noindent Some of the advantages of RBD-based reliability assessment are as follows \cite{goldberg1994system}:

\begin{itemize}

\item It provides an early assessment of the design concept by which changes in the design can be readily and economically incorporated.

\item An analyst can easily visualize the system design as compared to FTs.

%\item Components in an RBD configuration can be arranged in a manner to represent the component function in the system
\end{itemize}

\noindent Some of the limitations of this technique are as follows \cite{goldberg1994system}:

\begin{itemize}
\item Systems must be broken down into components, which requires significant amount of effort.

\item System element reliability estimates might not be readily available for all elements.

\item Not all systems can be modeled with combinations of series or parallel branches.
\end{itemize}

\subsubsection{Advantages and Limitations of Fault Trees}

\noindent Some of the advantages of FT based reliability evaluation are as follows \cite{goldberg1994system}:

\begin{itemize}
      \item Enables assessment of probabilities of combined faults/failures within a complex system.
      \item Single-point and common cause of failures can be identified and assessed.
      \item System vulnerabilities and low-payoff countermeasures are identified, which facilitates the
deployment of resources for improved control of risks.
\item Path sets can be used in trade studies to compare reduced failure probabilities with
increases in cost to implement countermeasures.
      \end{itemize}

\noindent Some of the limitations of this technique are as follows \cite{goldberg1994system}:

\begin{itemize}
      \item Allow addressing only one undesirable condition or event in one FT and thus usually require analyzing many FTs for the complete analysis of a particular system.
    \item FTs used for reliability assessment of large systems may not fit or run on conventional PC-based software. As the system increases in size and complexity, the corresponding FT also increases in size and requires more computational power \cite{chelson1971reliability,willie1978computer} for the generation of an accurate reliability assessment of the given system.
    \item A FT does not guarantee an accurate analysis unless all significant contributors of faults or failures are rightly anticipated.
\end{itemize}

\subsubsection{Advantages and Limitations of Markov Chains}

\noindent The advantages of using MC for analyzing reliability are as follows \cite{goldberg1994system}:

\begin{itemize}
        \item Provide a simple modeling approach for stochastic systems and an easy computation of the probability of an event resulting from the sequence of sub-events.
        \item System reconfiguration required by failures is easily incorporated in the model due to its simple modeling approach.
        \item Covered and uncovered failures of the components are usually mutually exclusive events and thus they cannot be easily modeled by using other techniques, such as RBD and FT, but are readily modeled by Markovian models.
        \end{itemize}

\noindent Some limitation of Markov modeling are as follows \cite{markovlimition}:

\begin{itemize}
        \item  Number of states increase exponentially as the size of the system increases. Markov state diagrams for such large systems are generally very complex and require computationally extensive manipulation using computer-based tools.
        \item Can only handle constant failure and repair rates, which limits its applicability for many real-world applications.
            \item Due to the memoryless property, the repair from a certain state may return the system to a new condition \cite{markovlimition}.
        \end{itemize}

\subsubsection{Advantages and Limitations of Bayesian Networks}

\noindent Some advantages utilizing BN modeling for reliability analysis are as follows:

\begin{itemize}
        \item  The graphical nature of a BN clearly displays the links between different system components, which allows better understanding of the system-component relationship and the effect of failure of components on the overall system.
        \item One obvious advantage of the BN over other reliability modeling techniques is that it uses prior information in order to estimate the reliability of new systems when very limited data about them is available.
        \end{itemize}

\noindent Some disadvantages of using BN modeling for reliability analysis are as follows:

\begin{itemize}
        \item  Sometimes prior information may not be accurate and thus leads to misleading conclusions.
        \item Just like Markov chains, this technique is unable to handle large and complex systems due to its state-based nature.
            \item It is often difficult to obtain the prior information of the system and calculate conditional probabilities by using prior data.
        \end{itemize}

\begin{table}
\caption{Comparison of Reliability Modeling Techniques}
\centering
\scriptsize
\scalebox{1}{
\begin{tabular}{|p{0.4\linewidth}|p{0.15\linewidth}|p{0.1\linewidth}|p{0.09\linewidth}|p{0.09\linewidth}|}
  \hline
   Features & Reliability Block Diagram & Fault Tree & Markov Chain & Bayesian Network\\
   \hline
  Success Domain & $\checkmark$  &  & $\checkmark$ & $\checkmark$\\
  \hline
    Failure Domain &  & $\checkmark$  & $\checkmark$ & $\checkmark$\\
    \hline
  Top Down Approach & $\checkmark$  & $\checkmark$  & $\checkmark$ & $\checkmark$\\
  \hline
 Identification and Prevention of Faults& $\checkmark$  & $\checkmark$  & $\checkmark$ & $\checkmark$\\
  \hline
  Combinatorial Problems& $\checkmark$  & $\checkmark$ & $\checkmark$ & $\checkmark$\\
  \hline
  Non-combinatorial Problems &  &  & $\checkmark$ & $\checkmark$\\
  \hline
 Large and Complex Systems & $\checkmark$  & $\checkmark$ & &\\
  \hline
\end{tabular}}\label{RFM_comp}
\end{table}

Based on the survey conducted in Section \ref{sec:depend_analysis}, we have found that the Markov Chain has been the most commonly utilized reliability modeling technique for communication networks. Moreover, the MC-based reliability models of the systems are frequently analyzed by all three analysis techniques, i.e., analytical, simulation and formal methods. On the other hand, utilization of RBD and FT models for the reliability analysis of communication networks is rapidly increasing specifically by using simulation tools. Other techniques, such as fuzzy logic and BN, have been also significantly contributing in the reliability analysis of communication networks domains in situations where the MC, RBD and FT have some limitations. The usage of formal methods based analysis of MC models of communication networks has also found some interest recently. On the other hand, to the best of our knowledge, the usage of FT and RBD models for the formal reliability analysis of communication networks is almost unexplored. We believe that this combination of modeling and analysis technique has a huge potential for ensuring accurate reliability analysis of safety-critical communication networks.
%Formal methods have been significantly contributing as an accurate alternative to these traditional reliability analysis techniques, especially in the saftey-critical applications of communication networks.

\subsection{Common Pitfalls of Reliability Modeling Techniques}
There are many challenges involved in evaluating and defining reliability for communication networks \cite{jereb1998network,bernardi2013dependability}. These challenges include (i) defining reliability adequately, (ii) the determination of possible states of the network, and (iii) the impact of failures on reliability for large number of network elements in the presence of multi-layer protection techniques.

RBD and FT can only be utilized when the communication network can be easily partitioned into logical blocks, such as multistage interconnection networks (MINs) \cite{bistouni2014pars}, and a rooted fault tree can be constructed based on the logical relationship of the network components failure, such as substation's communication network \cite{wang2013analysis}, respectively. Also, both these techniques have shown limitations when the behavior of the network is dynamic, i.e., the state of the system is changing with respect to time \cite{dantas,blake1989reliability}.

On the other hand, MC can capture the dynamic behaviour of the many complex communication networks, such as  randomized pulse modulation (RPM) \cite{ciufudean2007reliability}, and also overcome the limitations of RBD and FT by providing an extension to these techniques, such as DRDB and DFT. However, MC can only be applicable to those kinds of communication networks which can be modeled in the form of states and requires significant amount of memory and computational power to analyze large communication network systems \cite{markovlimition}. All three modeling techniques, i.e., RBD, FT and MC, require a statistical estimate of reliability of the sub-components of the network in order to evaluate the reliability of overall network system and are not applicable when the sub-component reliability related data is not available. This limitation in reliability estimation of communication networks has been handled by using the fuzzy logic technique, such as in single node WSN \cite{shu2008wireless} and communication control systems (CCS) \cite{peng2008approach}. BN utilizes the prior information to evaluate the communication network system reliability \cite{torres1998bayesian,wang2013analysis}. Thus, all these techniques are somewhat complementary in nature and have to used in the right context to evaluate the reliability of the given communication network.

\subsection{Comparison of Reliability Analysis Techniques}

In this section, we will compare and contrast the various techniques available for \textit{analyzing} reliability models. In particular, we will provide a discussion on the advantages and limitations of analytical, simulation, and formal-methods based reliability analysis techniques. A tabulated summary of this comparison is presented in Table \ref{table:comparison_tech}.

\subsubsection{Advantages and Limitations of Analytical Techniques}

Paper-and-pencil based analytical analysis is undoubtedly the most commonly used analysis technique for the reliability analysis of systems. The usage of mathematical modeling and reasoning ascertains accurate results. However, the involvement of manual manipulation and simplification, makes the analysis error-prone and the problem gets more severe while analyzing large systems. Moreover, it is possible---in fact a common occurrence---that many key assumptions required for the analytical proofs are in the mind of the mathematician and are not documented. These missing assumptions are thus not communicated to the design engineers and are ignored in the system implementations, which may also lead to erroneous designs. Moreover, paper-and-proof methods cannot be used to analyze the reliability of systems that have large models due to the manual nature of this analysis technique.

\subsubsection{Advantages and Limitations of Simulation Techniques}

The main strength of the simulation tools is their user friendly interface and analysis methods. Moreover, simulation is quite scalable as large models of systems can be easily manipulated via computers, which are very efficient in terms of book keeping. These benefits are usually attained by paying hefty licensing costs for the commercial tools and these tools usually require enormous computational resources. However, they cannot ensure absolute correctness as well due to the involvement of pseudo-random numbers and numerical methods and the inherent sampling-based nature of simulation.

\subsubsection{Advantages and Limitations of Formal Methods}

Due to the mathematical nature of the models and the involvement of mathematical and logical reasoning techniques in formal methods, the analysis results are guaranteed to be accurate and complete. However, these benefits are attained at the cost of heavy computational requirements, in the case of automatic theorem proving and model checking, and explicit proof guidance requirements, in the case of interactive theorem proving.

A comparison of the above-mentioned reliability analysis techniques is given in Table \ref{table:comparison_tech}. These techniques are evaluated according to their expressiveness, accuracy and the possibility of the automation of the analysis. Model checking and Petri Nets are not expressive enough to model and verify all sorts of reliability properties due to their state-based nature. The accuracy of the paper-and-pencil based proofs is questionable because they are prone to human errors. Simulation is inaccurate due to the involvement of pseudo-random number generators and computer arithmetics along with its inherent sampling-based nature. Theorem proving does not support all the reliability analysis foundations as of now. Finally, the paper-and-pencil based proof methods and interactive theorem proving based analysis involve human guidance and therefore are not categorized as automatic. However, there is some automatic verification support (e.g. \cite{slind2008brief}) available for theorem proving, which can ease the human interaction in proofs and thus we cannot consider interactive theorem proving as a completely manual approach. These days networking systems are extensively being used in many safety and financial critical applications, such as medicine, transportation and banking. Thus, the accuracy of their reliability analysis has become a dire need. As seen in Table \ref{table:comparison_tech}, only model checking, Petri Nets and theorem proving can fulfill these requirements, which makes formal methods a very interesting analysis technique for the analysis of safety-critical systems.

\begin{table}\caption{Comparison of Reliability Analysis Techniques}
\centering
\scriptsize
\scalebox{1}{
\begin{tabular}{|p{0.2\linewidth}|p{0.15\linewidth}|p{0.1\linewidth}|p{0.09\linewidth}|p{0.09\linewidth}|p{0.09\linewidth}|}
  \hline
  Feature & Paper-and-pencil Proof & Simulation Tools& Petri Nets &Theorem Proving &Model Checking \\
   \hline
%   \hline
  Expressiveness & $\checkmark$ & $\checkmark$& & $\checkmark$ &  \\
  \hline
  Accuracy & $\checkmark$&  & $\checkmark$ & $\checkmark$ & $\checkmark$ \\
  %\hline
%  Reliability-Availability analysis Support & $\checkmark$ & $\checkmark$ & $\checkmark$ & \xmark  & $\checkmark$ \\
  \hline
  Automation &  & $\checkmark$ & $\checkmark$ &   & $\checkmark$\\
  \hline
\end{tabular}}\label{table:comparison_tech}
\end{table}

\section{Conclusions}
\label{sec:conclusions}

In this paper, we have provided a tutorial introduction to the modeling and analysis techniques that have been used for studying reliability and availability of communication networks. We have discussed the various reliability models constructed using the building blocks offered by the formalisms of reliability block diagram, fault trees, Markov chains, and Bayesian network models. We have also presented background, and a critical comparison, of the various reliability analysis techniques (such as analytical methods, simulation modeling, and formal methods). Apart from providing the necessary background, we have also provided a detailed survey of the application of these techniques in the existing literature focused on studying reliability of communication networks. The main contribution of this work is that it is the first work that has presented a comprehensive review of the various techniques available for reliability modeling and analysis of communication networks along with a critical analysis describing their pros and cons in various contexts.

\section*{References}
 \bibliographystyle{elsarticle-num}
\bibliography{formal}
\end{document}